\documentclass[sigconf]{acmart}

\usepackage{enumitem}
\usepackage{amsmath}
\usepackage{color}

\AtBeginDocument{%
  }

\newcommand{\system}{ReSpark}

\begin{document}

\title{\system{}: Leveraging Previous Data Reports as References to Generate New Reports with LLMs}


\author{Yuan Tian}
\email{yuantian@zju.edu.cn}
\orcid{0009-0005-6089-7694}
\affiliation{
  \institution{Zhejiang University}
  \city{Hangzhou}
  \country{China}
}

\author{Chuhan Zhang}
\email{chuhan_z@outlook.com}
\orcid{0009-0000-8558-9579}
\affiliation{
  \institution{Zhejiang University}
  \city{Hangzhou}
  \country{China}
}

\author{Xiaotong Wang}
\email{xtitx327@gmail.com}
\orcid{0009-0006-2206-2749}
\affiliation{
  \institution{Zhejiang University}
  \city{Hangzhou}
  \country{China}
}

\author{Sitong Pan}
\email{pstongboa@gmail.com}
\orcid{0009-0006-7729-6497}
\affiliation{
  \institution{Zhejiang University}
  \city{Hangzhou}
  \country{China}
}

\author{Weiwei Cui}
\email{weiwei.cui@microsoft.com}
\orcid{0000-0003-0870-7628}
\affiliation{
  \institution{Microsoft}
  \city{Beijing}
  \country{China}
}

\author{Haidong Zhang}
\email{haizhang@microsoft.com}
\orcid{0000-0001-7411-8042}
\affiliation{
  \institution{Microsoft}
  \city{Beijing}
  \country{China}
}

\author{Dazhen Deng}
\authornote{Corresponding author: Dazhen Deng}
\email{dengdazhen@zju.edu.cn}
\orcid{0000-0002-9057-8353}
\affiliation{
  \institution{Zhejiang University}
  \city{Hangzhou}
  \country{China}
}

\author{Yingcai Wu}
\email{ycwu@zju.edu.cn}
\orcid{0000-0002-1119-3237}
\affiliation{
  \institution{Zhejiang University}
  \city{Hangzhou}
  \country{China}
}

\renewcommand{\shortauthors}{Tian et al.}

\begin{abstract}
Creating data reports is a labor-intensive task involving iterative data exploration, insight extraction, and narrative construction. A key challenge lies in composing the analysis logic-from defining objectives and transforming data to identifying and communicating insights. Manually crafting this logic can be cognitively demanding. While experienced analysts often reuse scripts from past projects, finding a perfect match for a new dataset is rare. Even when similar analyses are available online, they usually share only results or visualizations, not the underlying code, making reuse difficult. To address this, we present ReSpark, a system that leverages large language models (LLMs) to reverse-engineer analysis logic from existing reports and adapt it to new datasets. By generating draft analysis steps, ReSpark provides a warm start for users. It also supports interactive refinement, allowing users to inspect intermediate outputs, insert objectives, and revise content. We evaluate ReSpark through comparative and user studies, demonstrating its effectiveness in lowering the barrier to generating data reports without relying on existing analysis code.
\end{abstract}

\ccsdesc[500]{Human-centered computing~Visualization toolkits}
\ccsdesc[500]{Human-centered computing~Interactive systems and tools}

\keywords{Data Reports, Visualization Generation, Large Language Models}
\begin{teaserfigure}
  \includegraphics[width=\textwidth]{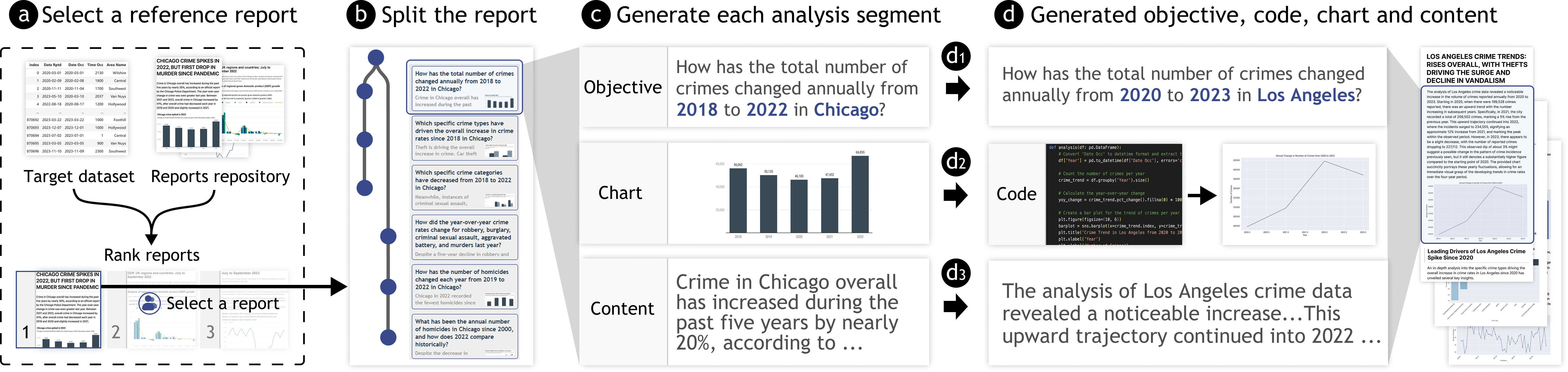}
  \caption{ReSpark Overview. ReSpark helps users generate new data reports by reusing existing ones.
It begins by ranking reports based on their alignment with the target dataset to assist in selecting a reference report (a).
The selected report is then segmented into a sequence of analysis segments (b).
For each segment, ReSpark extracts the analysis objective (c) and adapts it to the target dataset (d1).
It then generates the corresponding code and charts (d2), followed by textual insights (d3).
These elements are finally composed into a new report.}
  \Description{This figure illustrates the ReSpark workflow. (a) Users first view a dataset and select a reference report ranked by alignment. (b) The selected report is split into analysis segments. (c) Each segment contains an analytical objective, a chart, and a narrative. (d) ReSpark adapts each segment to the target dataset by modifying the objective (d1), generating new code and charts (d2), and rewriting the narrative (d3), which are composed into a new report.}
  \label{fig:teaser}
\end{teaserfigure}


\maketitle

\section{Introduction}

Data reports—typically composed of textual insights and expressive visualizations—are essential for communicating data value in business intelligence and industry.
Creating such reports is an iterative and cognitively demanding process, where data scientists transform data, configure charts, extract insights~\cite{behrens1997principles, subramanian2020tractus}, and organize them into a coherent narrative~\cite{li2023wherearewesofar}.
This process is time-consuming and mentally taxing~\cite{zhao2020chartseer, wongsuphasawat2017voyager, battle2019characterizing}.

To improve efficiency, reusing artifacts from previous data reports is a common strategy~\cite{epperson2022strategiesReuseTeam}.
These artifacts may include specific components, such as data transformation scripts or chart configurations, or higher-level ones, such as the analysis logic underlying the report.
Prior studies have explored recommending code snippets for specific data transformation and visualization tasks~\cite{li2023edassistant}.
With the rapid advancement of large language models (LLMs), code generation has become a popular approach to boost programming productivity.
However, as the length and complexity of the generated code increase, so does the likelihood of errors.
It remains difficult to generate an entire data report, including all the necessary code, in a single step.

More importantly, data reports are complex artifacts that require a high-level analysis logic to guide the LLMs.
Constructing and providing this logic remains cognitively expensive for users.
To address this, prior work on end-to-end data report generation can be broadly categorized into bottom-up and top-down approaches.
Most studies adopt a bottom-up perspective: Given a dataset, models heuristically search for candidate insights by combining different fields, which are then filtered and grouped into a report using optimization algorithms~\cite{shi2020calliope, sun2022erato, Socrates, dengDashBotInsightDrivenDashboard2022, HAIChart}.
While this method is efficient and mathematically elegant, it struggles to model high-level analysis logic that is built toward analytical goals, especially when the space of potential insights becomes large.
In contrast, top-down methods aim to construct an overall analysis logic that guides insight discovery.
For example, SNIL~\cite{cheng2024snil} analyzes the narrative structures of sports reports and introduces task-specific templates to guide LLMs, while LightVA~\cite{zhaoLightVALightweightVisual2024} decomposes high-level user intents into subtasks and generates corresponding code for multi-view visual analysis.
However, these top-down methods rely on predefined templates or LLMs' planning capability.
Human are deeply involved in constructing templates or controlling the LLM planning directions, which limits their scalability and generalizability.
 
In this study, we aim to improve both the efficiency and quality of generating analysis logic by reusing valuable structures from historical data reports.
To validate this idea, we conducted interviews with enterprise data scientists, which confirmed that report creation often relies heavily on prior experience and existing examples.
Analysts typically begin with a topic informed by data characteristics, domain knowledge, or analytical goals, and actively seek out similar past reports.
However, directly reusing previous reports poses practical challenges: most reports are shared only as final deliverables, without accompanying code or intermediate steps, making it difficult to replicate the analysis on new datasets.

We propose \system{}, an LLM-driven system that generates data reports by leveraging existing reports as references.
After uploading a dataset, users can either select a suitable report from our ranked candidates or upload a report of their own.
\system{} then uses this report to guide the generation of a new one.
The generated report is expected to follow the analysis logic of the reference while ensuring the correctness and relevance of insights applied to the new dataset.
The development of \system{} involves two major challenges:

\textbf{Reconstructing the missing analysis workflow.}
Most data reports lack explicit documentation of the analytical process, such as source code, transformation steps, or intermediate results.
Without access to these artifacts, it becomes difficult to infer the analytical objectives and operations that led to the final insights.
To address this challenge, we model analysis logic as a sequence of interdependent segments, each consisting of an analytical objective, a visualization, and a textual explanation.
We extract the dependencies between these segments and generate them sequentially to recover the logical progression of the original analysis.

\textbf{Making informed decisions on reuse and adaptation.}
Datasets often vary in collection time, source, format, and schema, making it non-trivial to directly reuse existing workflows. 
Reusing existing reports on new data may require reframing objectives or redefining metrics and comparisons.
However, deciding what to reuse and what to revise is often ambiguous and highly context-dependent.
To address this, we conducted a preliminary study comparing pairs of reports on similar topics to identify typical patterns and variations.
Building on these findings, we developed a pipeline that detects mismatches between the reference report and the new dataset, and generates adapted objectives, transformations, and insights.

Based on this method, we further developed an interactive interface for \system{} that allows users to insert new objectives, inspect real-time outputs, and revise generated content.
We evaluated \system{} through both comparative and usability studies, demonstrating its effectiveness in transforming reference reports into new, dataset-specific versions.

The contributions of this study are as follows:
\begin{itemize}[leftmargin=*]
    \item We propose a framework to generate new data reports by reusing past human-made reports to enhance the human-LLM collaboration for data report generation. 
    \item We identify the considerations for reusing data reports and summarize the required adaptation through a preliminary study. 
    \item We develop a proof-of-concept system for the framework, \system{}~\footnote{\url{https://github.com/ZJUIDG-AIVA/ReSpark}}, which can generate data reports by reusing existing reports and allow users to edit according to their preference. 
    \item We conduct comparative and usability studies to demonstrate the effectiveness of \system{}. The feedback could shed light on future research.
\end{itemize}
\section{Related Work}

In this section, we summarize the prior studies related to our work, encompassing authoring tools for data reports, reuse in data analysis, and LLM applications for data analysis.

\subsection{Authoring Tools for Data Reports}

Data reports leverage organized visualizations to convey data insights~\cite{segel2010narrative, masry2022chartqa, kang2021toonnote, FromDatatoStory}.
Prior work has outlined a typical workflow of authoring data reports~\cite{li2023wherearewesofar,li2023ai}, consisting of three stages: 
(1) \textit{analysis}, to explore the data and obtain data insights, 
(2) \textit{planning}, to prepare the core message and outline, and
(3) \textit{implementation}, to write text and create visualizations. 
This process demands expertise in both data analysis and insight organization, posing challenges for beginners to produce efficiently.

To address these challenges, researchers have developed various authoring tools~\cite{li2023wherearewesofar}. 
Some tools aim to reduce the manual workload through interactive interfaces~\cite{kim2019datatoon, lin2023inksight}.
For instance, DataToon~\cite{kim2019datatoon} supports pen and touch interactions that integrate analysis and presentation, enabling users to produce data comics more intuitively.
While these tools reduce the need for coding and context switching, they still rely heavily on users' decisions and manual interactions, which can be tedious and time-consuming.
Other tools use statistics-based algorithms to streamline the decision-making process~\cite{wang2019datashot,lu2021scrollytelling,shi2020calliope,shin2022roslingifier,chen2023calliopeNet, Socrates}.
For example, Calliope~\cite{shi2020calliope} automatically extracts insights and organizes them into structured reports.
Some systems target specific data types, such as time series~\cite{shin2022roslingifier} or network data~\cite{chen2023calliopeNet}. 
However, they overlook the semantic understanding of the data, which can limit the quality and relevance of the generated insights.
In this study, we use existing data reports as references to reconstruct their analytical workflows and apply them to new datasets using LLMs, incorporating the semantic richness to enhance the extraction and presentation of insights.

\subsection{LLMs for Data Analysis}

Recent advancements in large language models (LLMs), such as GPT-4~\cite{openai2023gpt4} and ChatGPT~\cite{chatgpt}, have demonstrated strong capabilities in semantic understanding and generation.
Driven by prompts, these models have been applied to diverse tasks, including code generation~\cite{wang2023dataFormulator}, information retrieval~\cite{suh2023Sensecape}, and insight extraction~\cite{ying2023livecharts, lin2023inksight}, motivating their use in data analysis scenarios.

A prominent direction is to generate visualizations from data and natural language~\cite{maddigan2023chat2vis}.
LIDA~\cite{dibia2023lida} and ChartGPT~\cite{tian2024chartgpt} decompose this task into several steps and process them sequentially, while Data Formulator~\cite{wang2023dataFormulator} focuses on generating transformation code to support visualization. 
Some studies further integrate visual outputs with narrative explanations~\cite{cheng2023gpt4analyst}.

More recent systems~\cite{ma2023insightpilot, zhaoLightVALightweightVisual2024, islam2024DataNarrativeAutomatedDataDriven, FromDatatoStory, hong2023aithreads} expand the use of LLMs to support iterative and continuous data analysis workflows. 
For example, LightVA~\cite{zhaoLightVALightweightVisual2024} decomposes high-level user intents into subtasks, generating data transformation and visualization code to build multi-view visual analyses.
These systems can produce structured multi-chart outputs with narrative elements, showing potential for generating data reports.
However, data report authoring remains a complex process driven by analytical logic.
Current tools struggle to generate comprehensive reports that fully reflect user intent based solely on an initial prompt.
Although some systems provide continuous interactive support~\cite{zhaoLightVALightweightVisual2024, hong2023aithreads} to help users refine the analysis logic, the process of communicating and articulating that logic remains cognitively demanding.

In this study, we introduce a reference-based approach.
Instead of relying solely on user-LLM interaction, we extract structured analysis logic from existing reports and adapt it to new datasets using LLMs.
Grounded in real examples, this approach leverages human-authored analytical knowledge to reduce the user's effort in constructing the analysis logic.

\subsection{Reuse in Data Analysis}
Several surveys have investigated various aspects of reusing scenarios in data analysis, particularly focusing on code reuse. 
Kery and Myers~\cite{kery2017exploring} noted that data scientists commonly reuse prior code versions during data exploration, and Koenzen et al.~\cite{koenzen2020code} noted the reuse of code from online external sources. 
Ritta et al.~\cite{ritta2022reusingMyOwnCode} studied the reuse patterns of expert data scientists and highlighted the frequent reuse of common abstraction codes such as package imports and visualizations. 
Additionally, Epperson et al.~\cite{epperson2022strategiesReuseTeam} discussed the strategies for code reuse in both personal and collaborative settings, identifying challenges such as the lack of modular code.
Other studies have explored the evolution of computational notebooks during reuse~\cite{liu2023refactoring, raghunandan2023codeEvolution}. 
Based on these surveys, researchers have developed tools to facilitate code reuse, including visualizing notebook changes~\cite{eckelt2024loops}, generating documents for notebooks~\cite{wang2022documentation}, and enabling parameter passing to notebooks~\cite{papermill}. 

However, these studies primarily focus on reusing functional code snippets, such as plotting graphs or importing packages, while overlooking the reuse of data analysis ideas and frameworks.
The studies most closely related to our approach are retrieve-then-adapt~\cite{qian2020retrieve} and MetaGlyph~\cite{ying2022metaglyph}, which reuse previous infographics as the basis for creating new ones.
Inspired by these works, we extend this idea to the reuse of existing data reports, aiming to adapt the full analytical process to new datasets.
\section{Problem Formulation}

In this section, we introduce the problem formulation of \system{}. 
First, we discuss our approach for generating data reports by breaking it down into three steps:
(1) selecting an existing report, 
(2) deducing a sequence of data analysis segments from the report, each comprising an analytical objective, data processing, and insights,
and (3) reproducing these segments on the target data. 
Based on the formulation, we conduct a preliminary study to investigate what aspects of existing reports can be reused and how they should be adjusted for new data.
The findings from this study inform the design considerations behind \system{}.

\subsection{Definition of Data Reports}
\label{subsec:problem_formulation}

Data reports are a type of data journalism that combines text and charts to convey insights from data. 
According to Hao et al.~\cite{designpatterns2025hao}, data journalism articles can be classified into five types: quick updates, briefings, chart descriptions, investigations, and in-depth investigations.
They form a spectrum along several dimensions: data source complexity, analytical depth, contextual richness, investigative effort, and discussion scope. 
From quick updates to in-depth investigations, articles gradually involve more complex data, deeper analysis, richer context, and greater reliance on human expertise.

Our work focuses on \textit{data reports} in the \textit{chart descriptions} category, which feature explicit narrative structures and multi-faceted analyses. 
Simpler types, such as quick updates and briefings, typically highlight isolated aspects of data dimensions. These reports are relatively stable in structure and do not rely on coherent narrative or analytical logic.
Therefore, they can often be generated using predefined templates~\cite{wang2019datashot}.
At the other end of the spectrum, investigations and in-depth investigations demand extensive incorporation of external knowledge beyond the data tables, making their automation highly challenging for current LLMs without significant human intervention.
In this study, we target chart description reports as a middle ground—more complex than simple briefs, yet still grounded in the data itself. These reports benefit from reusing prior analytical logic and domain expertise to guide dynamic report generation.
In our scenario, the intended users are data analysts in government or industry who frequently work with new datasets and are responsible for authoring such reports. Other than quick updates and briefings, they should discover insights and trends from the data and prepare more in-depth reports for the stakeholders.

To create such a data report, data analysts need to explore the data, obtain data insights, and organize them into coherent narratives and charts~\cite{li2023wherearewesofar}. 
A shortcut for this process might be referring to an existing report and adapting it with new data.
To begin, it is required to select a report relevant to the target dataset.
ReSpark provides a ranking mechanism to support this selection process, which we introduce later in Section~\ref{subsubsec:report_retrieval}.
In this section, we focus on the parsing and reconstruction of the selected report.

Given a report, the key to its reconstruction is to decompose the report into logically coherent segments and validate whether the new data supports the goals of each segment and yields similar insights. 
If not, how can adequate transformation of the original segments be employed to make the whole process successful? 

Previous research~\cite{bar2020automatically, li2023edassistant, batch2017interactive} has outlined the analysis workflow as an iterative process of three steps: 
(1) viewing the data and setting an analytical \textbf{objective},
(2) performing data \textbf{transformations} and encoding the transformed data into a chart~\cite{wang2023dataFormulator},
and (3) summarizing the \textbf{insights} from the results. 
We define each unit of this process as an ``\textbf{analysis segment},'' expressed as a triplet:
$$ segment := (objective, transformation, insight).$$
Data analysts typically construct reports by chaining such segments: deriving new objectives from previous insights and progressively generating follow-up segments.
Through this iterative process, they build a complete report grounded in the dataset.
In this study, we treat the segment as the basic building block of a data report.

\begin{figure}[!htb] 
  \centering
  \includegraphics[width=\columnwidth]{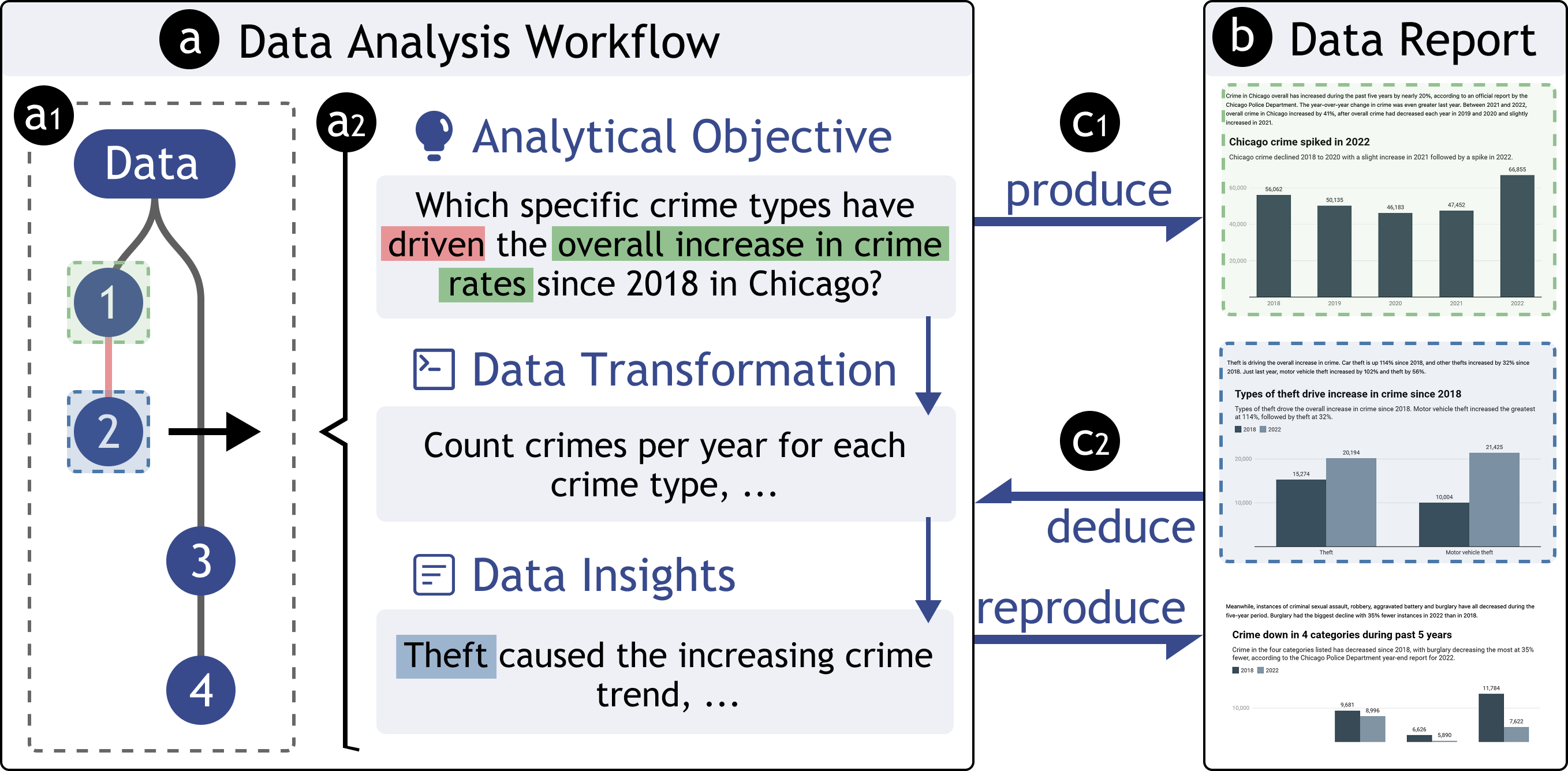}
  \caption{Producing a data report (c1) involves analyzing the data (a) and summarizing the analyzed insights into a data report (b). 
  Specifically, the data analysis workflow (a) includes a series of interdependent analysis segments (a1), each corresponding to an analytical objective, data transformations, and insights (a2). 
  To reuse an existing report on a new dataset, we first deduce its data analysis workflow and reproduce it on the new data (c2). }
  \Description{This figure illustrates how data reports are generated and reused. (a) The data analysis workflow begins with a dataset (a1) and proceeds through interdependent analysis segments, each consisting of an analytical objective, data transformations, and insights (a2). (b) These insights are organized into a data report. To reuse a report for new data, ReSpark first deduces the original analysis logic (c1) and then reproduces it with the target dataset (c2).}
  \label{fig:formulation}
\end{figure}

Formally, we model the analysis workflow as a sequence of interconnected segments $S = \{s_0, s_1, \cdots, s_N\}$.
Each segment $s_j = (o_j, t_j, i_j)$ comprises an analytical objective $o_j$, data transformations $t_j$, and insights $i_j$.
Dependencies among segments are defined as $D = \{d_0, d_1, \dots\}$, where each $d = (s_i, s_j)$ indicates that the objective of segment $s_j$ stems from the insights of $s_i$.
Specifically, the initial segment is rooted in the dataset.
Together, the analysis workflow, including $S$ and $D$, forms a tree structure (\autoref{fig:formulation}a1), where each node represents a segment and each edge denotes a dependency.
Under this formulation, a data report becomes a structured collection of insights distilled from the segment sequence $S$.

Our method centers on deducing and reproducing this sequence $S$ from the reference report, extracting and adjusting its objectives, transformations, and insights to fit the target dataset.
Finally, we organize the newly gleaned insights into a coherent report.

\subsection{Preliminary Study}
\label{subsec:preliminary_study_settings}

Based on the definitions above, our goal is to decompose a data report into a sequence of analysis segments and apply them to new datasets.
However, the flexible structure and narrative style of data reports can make it difficult to determine whether they can be consistently segmented.
Even if segmentation is possible, it remains unclear which components of these segments can be reused and which need to be adjusted for a new dataset.
To bridge this gap, we conducted a preliminary study with two objectives:
(1) to investigate whether the narrative structure of data reports aligns with identifiable analysis segments, and
(2) to identify the similarities and differences in analysis segments across reports on relevant topics.

We collected data reports of different topics from well-known organizations such as ONS~\cite{ons}, YouGov~\cite{yougov}, Pew Research Center~\cite{pewResearchCenter}, and PPIC~\cite{ppic}. 
To address the first objective, we analyzed the narrative structures of 35 reports to assess their alignment with analysis segments. 

For the second objective, we observed that these organizations often publish reports on similar datasets, such as periodic releases of epidemic data, typically issuing one report per dataset. 
These reports usually appear similar but vary subtly in analysis and insight content, which can be evidence to inspect which features can be inherited and which require adjustment. 

Therefore, we constructed 33 pairs of data reports that share the same topics and conducted pairwise comparisons to analyze their similarities and differences in terms of \textbf{analytical objectives}, \textbf{data transformations}, and \textbf{insight content}. 
Each report pair originates from the same organization and shares a common topic, often reflected in similar titles (e.g., Internet users, UK: 2018 and Internet users, UK: 2019), with variations in time or category (e.g., primary vs. secondary education). 
This pairing strategy allows us to examine how data reports are reused and adapted across different but related datasets.

A complete list of the reports, along with detailed statistics, is provided in the supplementary materials.
We present the findings for each of the two objectives in the following subsections.

\subsubsection{Narrative Structure of Reports}

To extract analysis segments from a report, we identified how the report's content aligns with the components of an analysis segment: \textbf{analytical objectives}, \textbf{data transformations}, and \textbf{insight content}.
We explored whether the report could be segmented so that each part corresponds to a distinct analysis segment.

Our analysis of 35 reports showed that in most cases (32/35), the report content was presented as distinct segments, each focused on a single objective rather than interspersed with multiple topics, with related text and possibly a chart grouped together.
Additionally, 11 reports included non-analytical content, such as background information, which could either supplement a specific analysis or the entire report and appeared flexibly throughout. 
Some reports (23/35) also included a summary of key insights at the beginning or end, which we excluded to focus on the main analytical content.

\subsubsection{Similarities and Differences in Pair Reports}

We analyzed patterns of similarity and difference between reports on the same topic to understand which components can be reused and which require adaptation. 
These findings inform our design of methods for reusing existing reports with new data.

\paragraph{Analytical objectives}
Analytical objectives are guidance for the exploration of insights from data.
For example, the report of internet users\footnote{\url{https://www.ons.gov.uk/businessindustryandtrade/itandinternetindustry/bulletins/internetusers/2018}} holds an objective of examining internet use across age groups.
Therefore, we identified the analytical objectives in the reports by inspecting the focus of their insights and compared them across 33 report pairs.

As a result, all of the pair reports reflect similar analytical objectives.
Specifically, 28/33 of them involve objectives that are exactly the same, while 29/33 of them show slight variations.
Most minor differences source from \textbf{variations in data context or scope}. 
For example, reports covering different time ranges often shifted their temporal focus accordingly.
Others result from \textbf{dependencies on earlier insights}. 
For example, a rising trend might trigger a follow-up analysis on its causes, so the changes in previous insights may cause adjustments in the latter objectives. 
Additionally, some pair reports involve completely different analytical objectives (15/33), which mainly stem from \textbf{different data fields}. 
For example, newer data may introduce additional data fields, thus triggering new objectives.
Reports may also incorporate insights from external data sources (4/33), broadening the scope or shifting the context, which also led to varied analytical objectives.

\paragraph{Transformation operations}
Data reports usually do not explicitly outline their data transformation steps. 
Moreover, 30/33 of our paired reports provide pre-processed datasets only, making it harder to infer the applied operations. 
Nonetheless, two facets of data processing can still be discerned:
(1) the outputs of transformation can be mirrored from the charts and narratives presenting data insights, and
(2) the underlying processing decisions can be reflected from the chart design choices, such as chart type and encoding.

Building on these observations, we compared content related to similar analytical objectives across each report pair, focusing on both the analysis output and the underlying transformation choices.
We observed that similar analytical objectives always yield similar \textbf{output forms}.
For example, trend analysis always results in a chart with time on the x-axis, indicating a transformed dataset aggregating values over time.
However, the \textbf{detailed processing choices} may vary to accommodate the data characteristics (12/33). 
Varied formats and scopes of data fields could potentially result in different chart types or levels of binning granularity to better align with visualization rules.

\paragraph{Insight Content}
The insight content, comprising both narratives and charts, is derived from the result of data processing. 
Since different datasets naturally produce different results, resulting in varying values in the content, our primary focus lies in the comparison beyond mere numerical distinctions. 

Regarding the similarity of insight content, we observe that each pair of reports shares a similar \textbf{narrative and visual style}, such as the formality degree in tone and infographic design. 
Since our primary goal is to reuse the analysis workflow, we do not consider the inheritance of content style in this study.
The differences primarily manifest in the textual descriptions of data insights. 
Different reports may describe varying types of data insights. 
For example, one report may highlight a notable outlier, while another describes overall trends.
These differences stem from \textbf{distinct analysis results}, which not only result in numerical disparities but also shape the nature of the insights described.
Even under identical transformation steps, one dataset may reveal a prominent outlier, while another may not.

\subsection{Expert Interview}
The findings of the preliminary study reveal that various aspects of existing data reports can be leveraged to generate new reports, but these elements require adjustments to align with the new data. 
The study also provides guidance on how to identify incompatible aspects and the directions in which adjustments should be made. However, it is unclear how users reuse past analysis reports for new scenarios.

Therefore, we conducted an expert interview targeted at data analysts who frequently explore new data and compose reports to communicate insights to leaders or clients. Moreover, we expected the interviewees to have certain experience in using LLMs in their analysis and prototyping process.
We interviewed two experienced data analysts, EA and EB. 
EA has over two years of experience working as an actuary at an insurance company, where they frequently analyze a variety of data, including claims history, policyholder demographics, risk factors, and market trends.
Findings are presented to a range of clients, including internal stakeholders (such as underwriters, product managers, and senior executives) and external clients (such as brokers, corporate policyholders, or regulatory bodies), to guide strategic decision-making and ensure compliance with industry standards. 
EB is an assistant professor in the department of journalism of a top-tier university. She also holds a Ph.D. in data science and frequently writes data journalism for news media.

We conducted 45-minute interviews with each expert, during which we raised the following questions: 
(1) What formats do they typically use to present data analysis results, such as slides, reports, and spreadsheets? 
(2) Do they encounter scenarios in which they reuse or refer to existing data analysis materials for analyzing new data and proposing new presentation materials? 
(3) If so, what is their workflow?

Both experts confirmed the use of data reports as a key format for communicating analytical findings, particularly in formal or structured settings. 
They also acknowledged the frequent reuse of past materials in these contexts. 
Their responses further revealed that reuse varies across formats, with different levels of effort needed. 
In particular, for data reports, they described concrete reuse considerations. 
We summarize these findings as follows:

\textbf{Data report is widely used, highly structured, and can be created by reusing existing materials.}
Both experts reported using data reports in formal scenarios and frequently referring to previous materials when composing new ones.
EA mentioned that he uses Excel files, slides, and data reports, and that for all three formats, he frequently refers to past materials. 
For instance, when analyzing data for a new insurance product, he often consults past reports for similar products. 
In such cases, both the data and the analysis objectives align closely, making existing materials particularly valuable. 
Additionally, EA noted that reports are his primary format for conveying data insights, particularly in formal settings.
EB explained that her approach to data presentation is scenario-dependent, and her use of previous materials varies accordingly. 
She highlighted that data reports, which are highly structured and commonly employed in formal contexts, often follow standardized templates. 
This structure makes it straightforward to follow existing materials when creating new reports.
EB further distinguished between such formal reports and more exploratory ``data stories,'' which, while similar to data reports in format, are more creative and exploratory. 
In these cases, she tends to focus on developing unique narratives that reflect individual insights and perspectives.
This contrast suggests that the reuse potential is influenced by the communicative goals and format.
In this study, we focus on data reports, which are reasonable to reuse existing materials. 

\textbf{Reuse challenges differ across formats and present unique potential for reports.}
EA emphasized that reuse workflows differ by format.
For Excel files, past documents often contain reusable queries and formulas and can be updated with simple changes.
In contrast, slides and data reports typically lack the original code. While prior examples can help shape analysis goals and insight writing, the analysis must be redone manually.
EA also distinguished between the two: slides are often visual and presentation-oriented, lacking narrative descriptions, whereas data reports contain detailed textual findings.
This self-contained nature provides potential for identifying reusable analysis logic from data reports.

\textbf{Reuse and adjustment depend on data similarity and evolving insights.}
EA and EB further elaborated on the workflow of reusing previous data reports. 
When the data fields and analytical objectives are similar, both experts noted that the overall structure of the report can be reused directly, requiring only chart replacements and updates to numerical conclusions.
However, data field differences require more substantial adjustments, such as modifying the objectives and even developing new directions.
In cases where the insights change, the associated insight content also needs to be completely rewritten.
These observations echo the findings of our preliminary study and further validate the need for systematic support when reusing reports on new datasets.

\subsection{Design Considerations}

Based on the problem formulation, preliminary study, and expert interview, we summarize five design considerations (\textbf{C1}-\textbf{C5}) for developing an automated method to reuse existing data reports with new datasets.

\begin{enumerate} [leftmargin=*, label=\textbf{C\arabic*}]
\item \textbf{Support analytical objective extraction, correction, and addition. }
The core of the method is to extract and re-execute the analysis workflow, represented as a sequence of analytical objectives.
To achieve this, the system should first segment the report into distinct analysis segments, then extract their objectives and their dependencies.
It should further support correcting or adding objectives based on inter-segment logic and data features.

\item \textbf{Generate appropriate data processing steps automatically. } 
Based on the preliminary study, data processing steps are often not explicitly documented in reports.
The method should be capable of inferring transformations from the original report while also adapting to the new dataset's characteristics.

\item \textbf{Produce informative insight content from analysis. } 
Charts and narratives are the main components of data reports.
The system should generate insight content that communicates new data insights based on analysis results.
Our preliminary study also revealed that reports often include non-analytical content (e.g., background information).
As LLMs are trained on broad external knowledge, they can generate such content, but it should be marked, as it may not always be reliable.

\item \textbf{Enable real-time output observation and interactive editing. } 
Given the complexity of the method, which involves analytical objectives, data processing, and insight content, uncertainties may arise that deviate from user expectations.
The system should provide an interactive interface that supports real-time inspection and modification of outputs to ensure alignment with user needs.

\item \textbf{Facilitate structural organization and refinement.}
Since the generated report is based on the reference report's analysis logic, it will naturally exhibit a similar narrative structure. 
The system should support reorganizing this structure, including generating or modifying (sub-)titles and flexibly arranging sections, to better fit the new content. 

\end{enumerate}

Based on these considerations, we develop \system{}, an intelligent system to deduce and reproduce the authoring workflow of existing reports on new data. 
The pipeline consists of three stages:
\textbf{In the pre-processing stage,} \system{} ranks the reports in its repository by similarity and relevance to the user's dataset, from which the user can select a suitable reference.
It then decomposes the selected report into interconnected segments and extracts the analytical objectives of each segment (\textbf{C1}). 
\textbf{In the analysis stage,} \system{} generates each segment by reusing information from the reference report. 
It identifies inconsistencies and customizes objectives, transformations, and insights based on the new data (\textbf{C1-C3}). 
\textbf{In the organization stage,} \system{} inherits the original report structure and enables title re-generation (\textbf{C5}). 
An interactive interface allows users to monitor intermediate results, add new objectives, and edit content as needed (\textbf{C4}).
\section{\system{}}
\label{sec:respark}

This section describes the implementation of \system{}, including three stages: pre-processing, analysis, and organization. 
All stages are powered by LLMs, which drive the core reasoning and generation processes.
The prompts used for each stage are provided in the supplementary materials.

\subsection{Pre-processing Stage}

Before proceeding with analysis and organization, we pre-process the dataset and reports to (1) acquire necessary data features, (2) rank reports based on the target data, and (3) extract the analytical objectives and their dependencies of the selected report for subsequent tasks. 

\subsubsection{Data Summarization}
\label{subsubsec:data_pre_processing}

Based on the preliminary study, most adjustments entail considerations of data features such as context, scope, fields, and formats. 
Presenting the entire dataset to LLMs is impractical due to token limitations, and it does not facilitate a comprehensive understanding of these data features. 
Therefore, we utilize a similar data summary method to LIDA~\cite{dibia2023lida}. 
This method first extracts scope, data type, and unique value count information and samples some values for each data field. 
Subsequently, it utilizes LLMs to provide brief semantic descriptions for the dataset and each data field. 
The description of the dataset can also help rank relevant reports. 
These pieces of information are then integrated to form a data summary.

\subsubsection{Report Segmentation}
\label{subsubsec:report_pre_processing}

Based on the problem formulation, our goal is to deduce the analysis segments and their dependencies from the reference report for subsequent execution. 
Our preliminary study revealed that most reports present analytical content in well-separated segments, each centered on a single objective with related text and visuals grouped together.
This suggests that, ideally, we can find a segmentation that aligns each section with an analysis segment. 
In this light, \system{} should segment the report accordingly, extract the analytical objectives for each segment, and deduce their dependencies.

To achieve this goal, an appropriate report segmentation criterion is very important, as it directly determines the entire analysis workflow. 
The accuracy of segmentation also affects the quality of the extracted analytical objectives and dependencies.

We were initially inspired by prior work in automatic storytelling and insight mining, which formalizes data stories as sequences of structured insights~\cite{ma2023insightpilot, wang2019datashot}.
For example, Calliope~\cite{shi2020calliope} defines an insight as a quantified pattern within a data subspace based on a measurable field.
This formulation suggests that reports could be segmented by identifying the insight type, measure, subset, etc., and combining the insights into segments based on these labels. 
However, we found this approach difficult to apply in practice, as real-world reports often involve more complex analysis, including multi-step transformations and reasoning over derived variables.

Therefore, we need to define new segmentation criteria that accommodate the flexible analysis in the report. 
Instead of formally defining a ``segment'' or an ``analytical objective'' with a strict data model, we provide a loose description of how segments can be divided and use LLMs to perform segmentation. 
Guided by our preliminary study, which showed that most analytical segments are composed of contiguous text and optionally a chart, we segment the report, extract analytical objectives, and infer inter-segment dependencies using the following approach:

\begin{itemize}[leftmargin=*]
    \item \textbf{Match. } 
    For each chart, we match the related paragraphs to form a segment. 
    Based on our preliminary study, we assume that (1) each paragraph corresponds to the nearest preceding or following chart, or to none at all, and (2) all text associated with a chart appears as a continuous block. 
    We process the report paragraph by paragraph, prompting LLMs to determine whether each paragraph describes insights from the nearest chart and should be linked to it.
    
    \item \textbf{Categorize. } 
    For paragraphs not matched with any chart, LLMs categorize them to determine if they are analytical (e.g., discussing data findings) or non-analytical (e.g., background information).
    For analytical text blocks, LLMs further evaluate whether they constitute a single segment, based on whether the content derives from the same transformed data.
    
    \item \textbf{Summarize. } 
    After matching and categorizing, LLMs summarize the analytical objective of each segment and deduce its dependencies with previous segments. 
    Following Calliope~\cite{shi2020calliope}, we adopt six types of logical relations (e.g., comparison, elaboration) to express inter-segment dependencies.
    LLMs determine whether a segment builds upon a previous one or originates independently from the data.
\end{itemize}

We evaluated the performance of our segmentation method using F1 score, precision, and recall.
Specifically, we randomly sampled 10 reports from the 35 analyzed in our preliminary study.
Three authors independently annotated the segment boundaries and resolved disagreements through discussion to establish the ground truth.
We then applied our segmentation method to the same reports and computed three metrics by comparing the predicted segment boundaries with the annotated ground truth.
Notably, our method is model-agnostic; we tested it with both GPT-4o and Qwen2.5-VL-72B-Instruct (deployed on Ascend 910b), respectively.

Using GPT-4o, the method achieved an F1 score of 90.9\% (88.5\% precision, 93.4\% recall), indicating strong alignment with human annotations and demonstrating the effectiveness of our approach.
Qwen2.5-VL also produced comparable results, with an F1 score of 90.8\% (89.4\% precision, 92.3\% recall), suggesting that the method generalizes well across different models.
Given the slightly higher F1 score, we chose GPT-4o as the default model in our comparative and usability studies.


\begin{figure*}[!htb] 
  \centering
  \includegraphics[width=\textwidth]{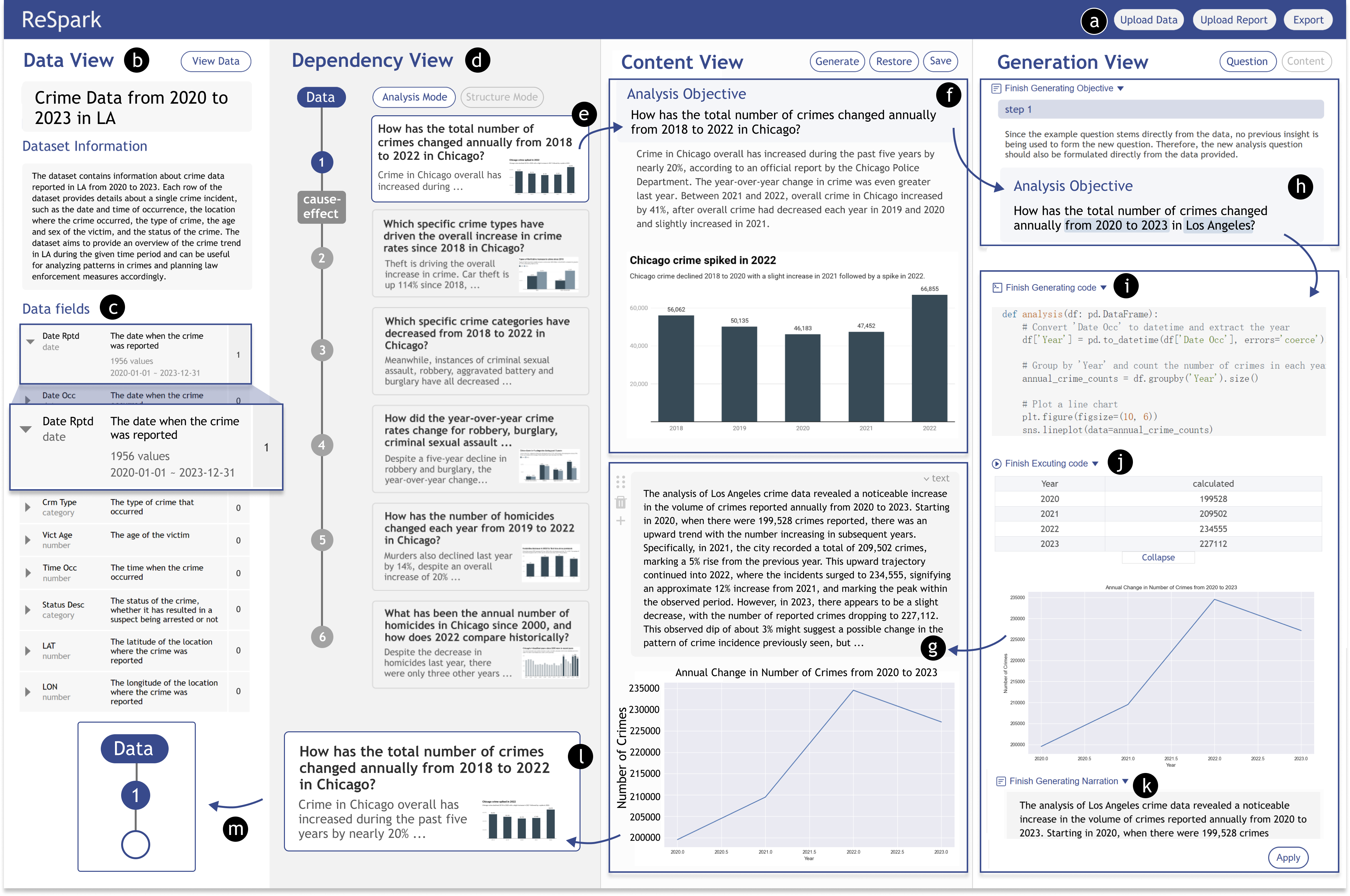}
  \caption{
  The interface of \system{}. 
  \system{} consists of four views: data view (b-c), dependency view (d-e), content view (f-g), and generation view (h-k). 
  The data view displays the dataset description and field information. 
  The dependency view displays the extracted interdependent analysis segments. 
  The content view shows the analytical objective and content of the selected segment. 
  The generation view shows real-time generated results. 
  }
  \Description{
  This figure shows the interface of ReSpark, consisting of four main views: (b–c) Data View displays the dataset’s description and detailed data fields. (d–e) Dependency View presents the extracted analysis segments and their dependencies. (f–g) Content View shows the selected segment's analytical objective, chart, and narrative. (h–k) Generation View supports generating and editing new segments based on the target dataset, including defining new objectives (h), generating code (i), executing results (j), and producing narratives (k). The system enables users to trace, inspect, and adapt analysis logic interactively.
  }
  \label{fig:interface}
\end{figure*}

\subsubsection{Report Ranking}
\label{subsubsec:report_retrieval}

With the summarized data and segmented reports, users need to select a suitable reference report to adapt for their target dataset.
As shown in our preliminary study, various aspects of existing reports, such as analytical objectives and insight content, can provide useful guidance. 
However, differences in the underlying datasets mean that certain parts of the reference report may require modification.
In general, the more similar the reference report's dataset is to the target dataset, the more seamlessly its content can be reused.
Thus, identifying such suitable reference reports is essential for effective reuse.

To assist users in selecting reports, \system{} ranks candidate reports by their dataset similarity to the target dataset.
The key idea is to embed relevant information from both the dataset and each candidate report into vector representations, compute their cosine similarity, and sort the reports accordingly.
A central challenge lies in deciding which information from the dataset and report should be embedded.
We propose two complementary mechanisms: topic relevance and field similarity.

\textbf{Topic relevance} captures the semantic alignment between the general domains of the dataset and the report.
For example, a dataset on consumer spending is topically aligned with reports about market trends.
To compute topic relevance, we embed the dataset's name and description, and for each report, embed its section headings and extracted analytical objectives.
We assume that these elements reflect the report's high-level domain, and that the cosine similarity between their embeddings serves as a reliable proxy for topic alignment.

\textbf{Field similarity} captures the correspondence between the data fields in the report and those in the target dataset.  
For instance, a report that discusses gender-based voting trends is more aligned with a dataset containing gender and demographic information.  
To compute field similarity, we embed the names and descriptions of the target dataset's data fields.  
For each report, we prompt LLMs to infer the key fields involved in the analysis, and embed both the inferred field names and their descriptions. 
The cosine similarity between these embeddings indicates their field-level alignment.

We sum the topic relevance and field similarity scores for each report to produce a final ranking.
The ranked list is presented to the user, who can then select the most suitable reference report to guide the subsequent analysis.

\subsection{Analysis Stage}
\label{subsec:analysis_stage}

After the pre-processing stage, we obtain the target dataset summary, the selected reference report, and its segmented structure.
Each segment includes an analytical objective, its dependency, and associated content such as text and charts.
The next stage is to reproduce the analysis workflow by generating each segment on the new data, encompassing reusing and reconstructing the analytical objective, data transformations, and insight content. 

\subsubsection{Analytical Objective Correction and Insertion}

The analysis workflow is driven by the sequence of posed analytical objectives. 
While these objectives are extracted from the reference report, they often require adjustment or extension to accommodate differences in the target dataset.
\system{} supports two key capabilities: (1) correcting analytical objectives to align with the target dataset and (2) inserting new objectives according to the data features and segment dependencies. 

\textbf{Objective correction. }
Our preliminary study revealed that while analytical objectives from reference reports are often reusable, they frequently require adjustment due to differences in data fields, context, or logical dependencies. 
To address these issues, \system{} applies LLM-based correction along three primary dimensions.

\textit{Data fields. }
The extracted analytical objectives may reference data fields that do not exactly match the target dataset.
Given the pre-processed field summary, we prompt LLMs to assess whether the target dataset contains fields that are semantically aligned with those mentioned in the objective, even when names differ (e.g., relating ``earn money'' to ``gross'').
If alignment is found, the model explains its rationale and outlines the transformations needed to fulfill the objective.
If no adequate substitute exists, the model identifies which external fields are missing and suggests available data fields as alternatives.
For instance, if a movie dataset lacks geographic data, the objective of locating the highest-grossing movies might shift focus to their directors.

\textit{Insight dependency.}
Some objectives depend on insights from earlier segments.
When these upstream insights shift (e.g., from increasing to decreasing trends), the dependent objectives must also be revised (e.g., from explaining growth to analyzing decline).
Additionally, if the dependency implies a broader scope or context (e.g., moving from local to national trends) that the target dataset does not support, the objective may need to be reframed or removed.
To address this, we provide LLMs with the updated results of prior segments and the dataset summary, prompting them to assess whether the dependency remains valid and how the objective should be adjusted.

\textit{Data scope.}
Third, minor adjustments are often required for differences in data context or coverage (e.g., time range or population).
For example, an objective focused on a five-year trend needs revision if the dataset only spans three years.
LLMs revise such objectives to match the contextual scope of the available data.

\textbf{Objective insertion. }
In addition to correcting existing objectives, users may also want to explore new directions or analyze additional data fields. 
To support this, \system{} provides an objective insertion feature that allows users to specify one or more data fields of interest and define the logical relationship to a previous segment—selected from six predefined analytical logics, such as similarity or contrast~\cite{shi2020calliope}.
Based on this input, the system uses LLMs to generate a candidate objective for insertion.

\subsubsection{Data Transformation Generation}

Once the analytical objective is refined, \system{} proceeds to generate the corresponding data transformation code.
Since these operations are not explicitly stated in the report, we utilize the code-generation capabilities of LLMs for this phase. 

LLMs are prompted to reason through the steps required to fulfill the objective, drawing inspiration from the reference report while adapting to the new dataset.
We instruct the model to first outline its plan (i.e., using step-by-step reasoning~\cite{kojima2022large}), and then produce Python code that transforms the data and visualizes it using libraries like Matplotlib~\cite{Hunter2007matplotlib} or Seaborn~\cite{Waskom2021seaborn}.
To guide LLMs in generating appropriate charts, we include prompt instructions such as ``use the right chart type,'' and ``consider the data field types.'' 
\system{} executes the generated code and obtains the transformed data table and chart.
Any runtime errors or unexpected results are returned to the LLMs for iterative refinement, and the process iterates until successful execution.

\subsubsection{Insight Content Generation}
Once the transformation results are obtained, \system{} generates the corresponding textual insight.
LLMs are prompted to describe the patterns observed in the transformed data and chart, producing clear and informative narratives.
To support generation, the system also provides the insight content from the reference report as contextual reference, helping the model understand how data insights were described.

\subsubsection{User Customization}
After generating a segment, users can view the adapted objective, generated code, resulting data table and chart, and corresponding insight text.
\system{} supports customization by allowing users to modify the objective (e.g., adding instructions such as ``use a bar chart'') and regenerate the analysis.
This enables users to better align the analysis method, chart design, and textual insights with their intent. 
Users can also directly edit the generated code and text to produce the desired results.

\subsection{Organization Stage}

After reproducing the analysis workflow and generating new data insights, the final step is to organize the results into a coherent report.
\system{} inherits the structural layout of the reference report by default, including section groupings and titles, and incorporates newly added segments at appropriate positions based on their dependencies.
Therefore, the resulting report naturally follows the original report's implicit logical structure.

To support flexible editing, users can regroup segments into custom sections and trigger title regeneration using LLMs, which generate new titles and headings based on the updated content.
Alternatively, users may manually revise titles as needed, allowing the report to better reflect their preferences.

\section{Interface}
\label{sec:usage_scenario}

We incorporate \system{} with an interactive interface. 
The interface consists of four views: data view, dependency view, content view, and generation view. 
We introduce it through a usage scenario, where we use GPT-4o as the foundation model.

\textbf{Pre-processing. }
A data analyst is tasked with creating a data report for a dataset detailing crime in Los Angeles from 2020 to 2023~\footnote{\url{https://catalog.data.gov/dataset/crime-data-from-2020-to-present}}. 
S/he decides to use \system{} to assist in creating a data report with the Los Angeles data. 
The analyst begins by uploading the Los Angeles crime dataset (\autoref{fig:interface}a). 
\system{} then pre-process the dataset. 
After the pre-processing, the analyst can view the dataset information (\autoref{fig:interface}b), including the file name, overall description, and data field information. 

\begin{figure}[!htb] 
  \centering
  \includegraphics[width=\columnwidth]{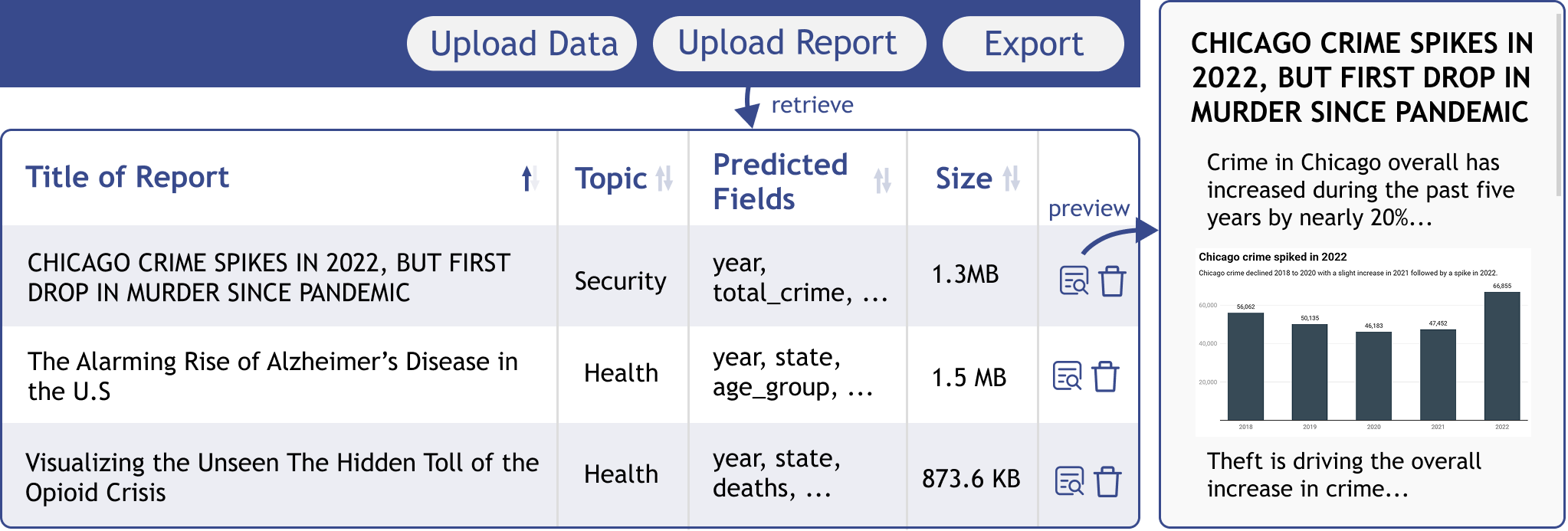}
  \caption{
    A demonstration of the ranked reports and their summarized information.
  }
  \Description{
  The figure shows the interface for selecting a reference report in ReSpark. On the left, a ranked list displays available reports along with their titles, topics, predicted data fields, and file sizes. Users can preview report content or remove reports using action buttons. On the right, a preview panel shows the selected report's title, summary text, and an example chart, helping users assess its relevance before selection.
  }
  \label{fig:retrieve}
\end{figure}

Next, the analyst opens the report repository (\autoref{fig:interface}a), which displays the available reference reports ranked by alignment to the dataset (\autoref{fig:retrieve}). 
The analyst selects the highest-ranked report, a Chicago crime report~\footnote{\url{https://www.illinoispolicy.org/chicago-crime-spikes-in-2022-but-first-drop-in-murder-since-pandemic/}}. 
The report is divided into six interdependent segments, each corresponding to an analytical objective and report content (\autoref{fig:interface}d).
With the segmentation and extracted objectives, the analyst can proceed with generating a new report.

\textbf{Segment generation. }
The analyst starts with the first segment (\autoref{fig:interface}e), focusing on crime trends in Chicago from 2018 to 2022. 
S/he clicks ``generate'' at the top of the content view to adapt this analytical objective to the Los Angeles data. 
\system{} modifies the objective to ``How has the total number of crimes changed annually from 2020 to 2023 in Los Angeles?'', which is aligned with the context and scope of the Los Angeles data (\autoref{fig:interface}h). 
Following this, the analyst clicks ``generate'' again to conduct the analysis. 
\system{} generates the code and executes it (\autoref{fig:interface}i), obtains the transformed data and chart, and generates the narratives to describe the findings. 
The resulting chart shows the number of crimes and changed percentages each year (\autoref{fig:interface}j). 
The narrative describes the overall increase and the year-by-year changes, especially the peak in 2022 (\autoref{fig:interface}k). 
The analyst considers such results reasonable and applies them (\autoref{fig:interface}g).

\begin{figure}[!htb] 
  \centering
  \includegraphics[width=\columnwidth]{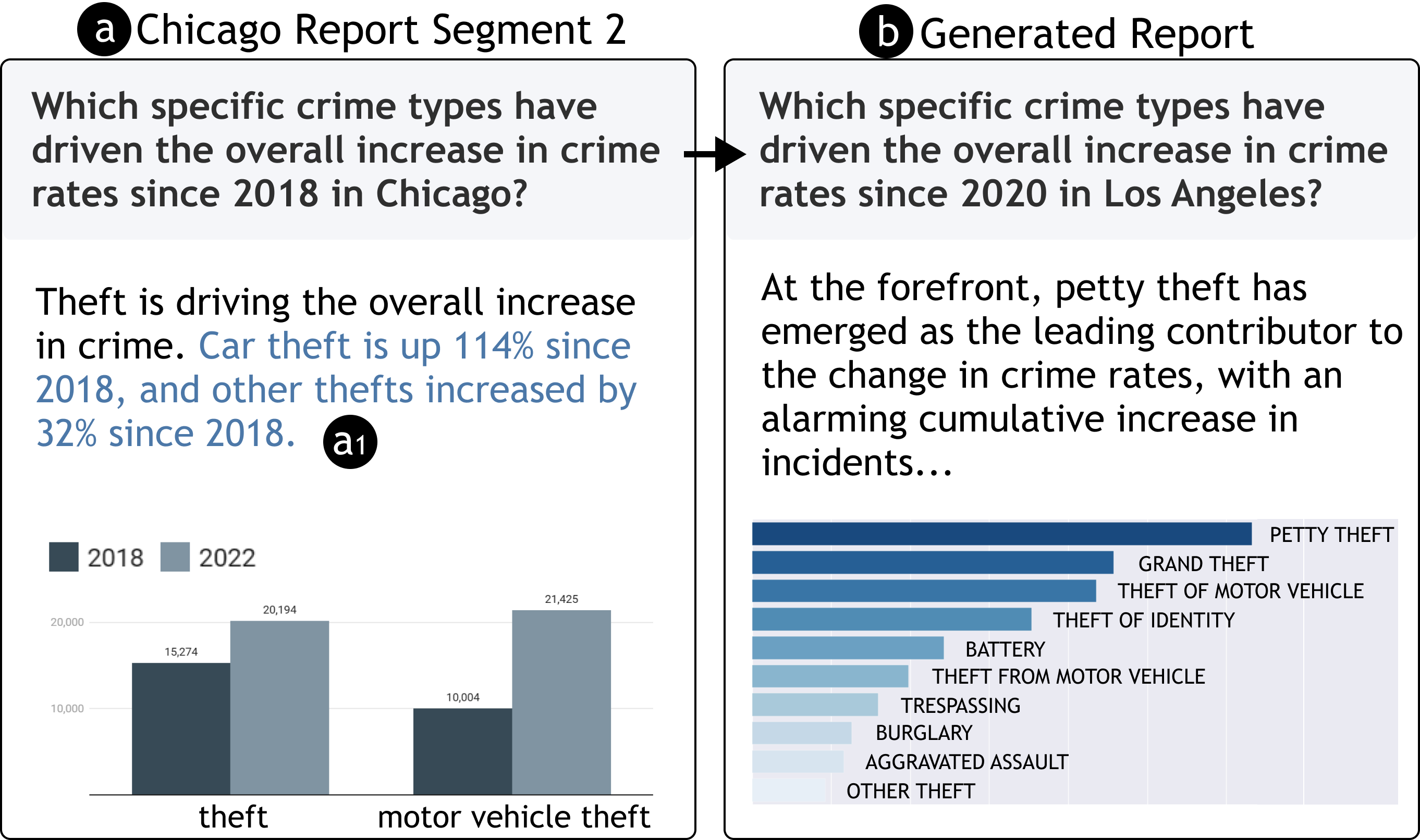}
  \caption{
    The process of generating the second segment. 
  }
  \Description{
  An illustration of how ReSpark generates the second segment by reusing a reference. (a) shows the second segment from the original Chicago report, analyzing which crime types drove the overall increase. It includes an objective, a textual insight highlighting theft and car theft, and a bar chart. (b) shows the corresponding segment in the generated Los Angeles report, adapted for a different time range and dataset. It reuses the analytical structure but updates the content to reflect new leading crime types such as petty theft.
  }
  \label{fig:segment_2}
\end{figure}

The second segment aims to explore the crime types that drove the increasing trend, which depends on the first segment with a cause-effect logic (\autoref{fig:interface}d). 
As the previous segment results in an overall increasing trend as well, \system{} inherits the logic and corrects the objective to match the Los Angeles data context (\autoref{fig:segment_2}b). 
Different from the reference report which only shows the changes in 2018 and 2022, \system{} calculates the top ten crime types contributing to the uptick and visualizes them through a bar chart for a clear comparison of cumulative numbers. 
Such varied design choices may stem from the narrative description in the Chicago report, which describes both the overall change and last year's change (\autoref{fig:segment_2}a1). 
As the analysis progresses, the analyst explores the subsequent two segments, including the decreased crime types and their changes from 2022 to 2023. 

\begin{figure}[!htb] 
  \centering
  \includegraphics[width=\columnwidth]{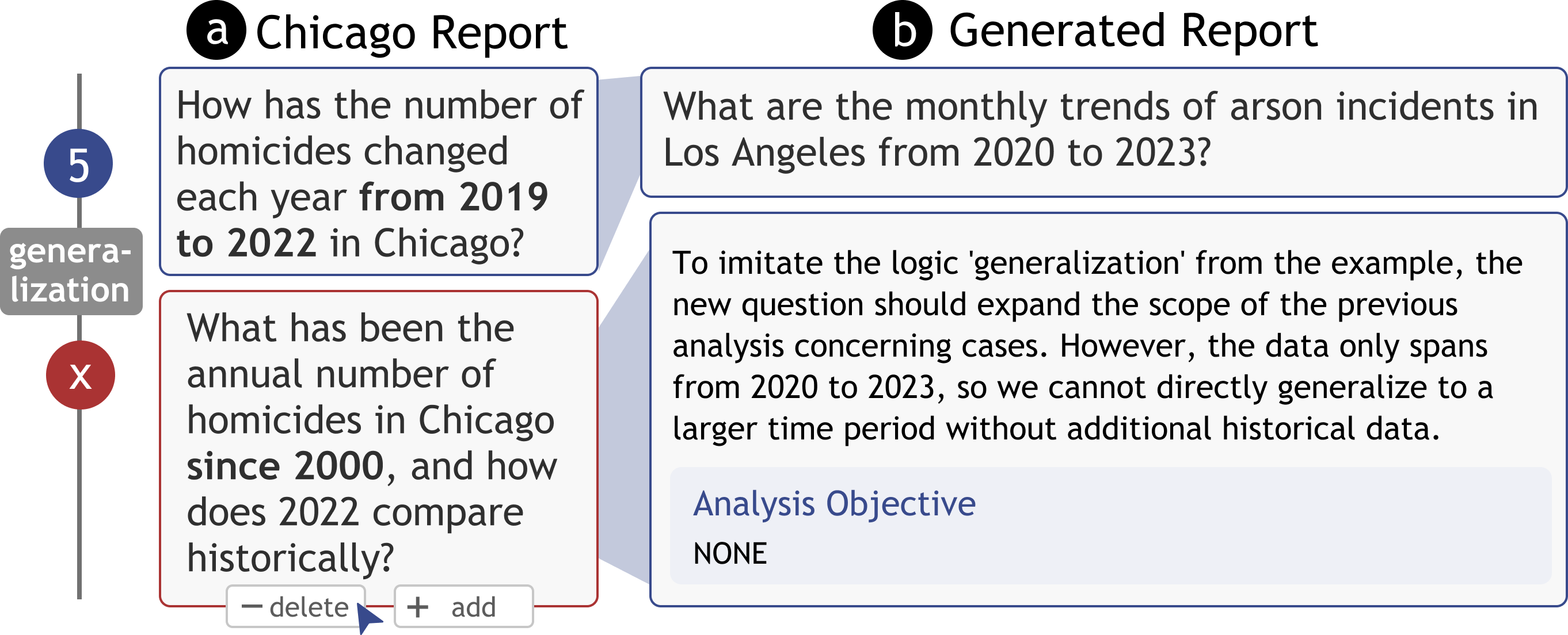}
  \caption{
  A demonstration of an analytical objective that fails to be corrected and needs to be removed. 
  }
  \Description{
  This figure shows a segment that needs to be removed in ReSpark. (a) The original objective asks about the change in homicides from 2019 to 2022 in Chicago. The rewritten version attempts to generalize it to a longer historical period (since 2000), which the target dataset does not support. (b) ReSpark identifies that the target dataset only spans from 2020 to 2023 and therefore cannot answer the generalized question, marking the objective as invalid and suggesting its removal.
  }
  \label{fig:delete}
\end{figure}

\textbf{Objective removal. }
After the four segments above, the Chicago report moves to another analytical objective on homicide trends from 2019 to 2022 (\autoref{fig:delete}). 
Similarly, the analyst applies the tailored objective and report content generated by \system{}, which analyzes the trend of arson in LA from 2020 to 2023. 
However, the next segment of the Chicago report generalizes its analysis to include homicide trends from 2000 to 2022. 
To inherit such a generalization logic on the generated report with the LA data, external data sources are needed, as the provided dataset covers only 2020 to 2023. 
Identified such a case, \system{} marks the analytical objective as ``none'' and visually indicates this error through a red node. 
Consequently, the analyst removes this segment.

\begin{figure}[!htb] 
  \centering
  \includegraphics[width=\columnwidth]{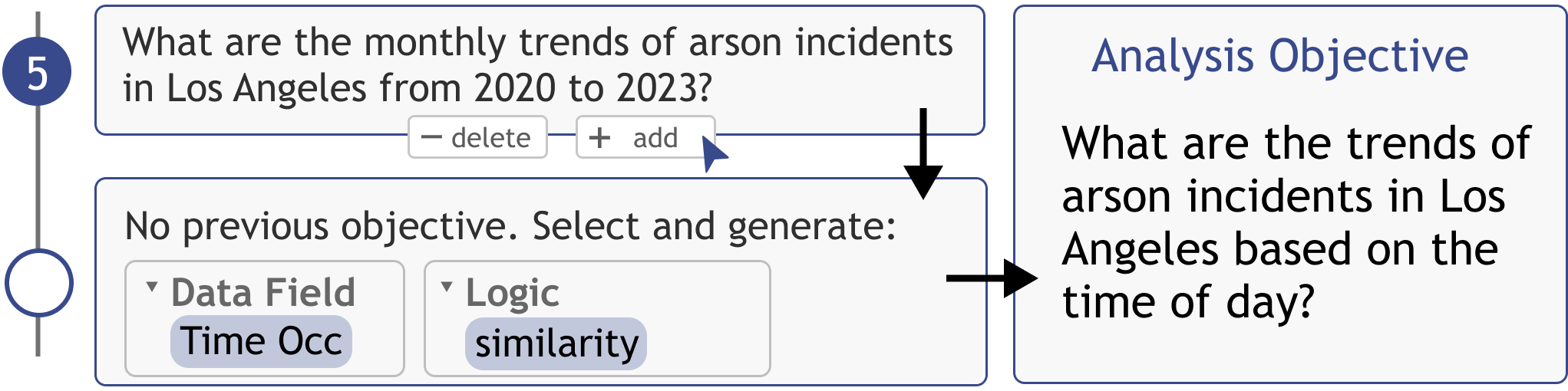}
  \caption{
  A demonstration of inserting an analysis segment and generating a new analytical objective. 
  }
  \Description{
  This figure shows how a user inserts a new analysis segment in ReSpark. The user first adds a new segment and selects a relevant data field (``Time Occ'') and logic pattern (``similarity''). ReSpark then generates a new analytical objective: ``What are the trends of arson incidents in Los Angeles based on the time of day?''
  }
  \label{fig:add}
\end{figure}

\textbf{Objective insertion. }
After generating the segments, the analyst refers to the list of data fields (\autoref{fig:interface}c), where the rightmost column displays the usage count of each field. 
Noticing that the time-related field ``Time Occ'' is under-analyzed, the analyst decides to insert a new node focusing on time, building on the previous analysis. 
S/he selects the ``Time Occ'' field and applies a ``similarity'' logic to the previous segment for parallel analysis (\autoref{fig:add}). 
\system{} then forms a new analytical objective to explore the time distribution of arson.

\begin{figure}[!htb] 
  \centering
  \includegraphics[width=\columnwidth]{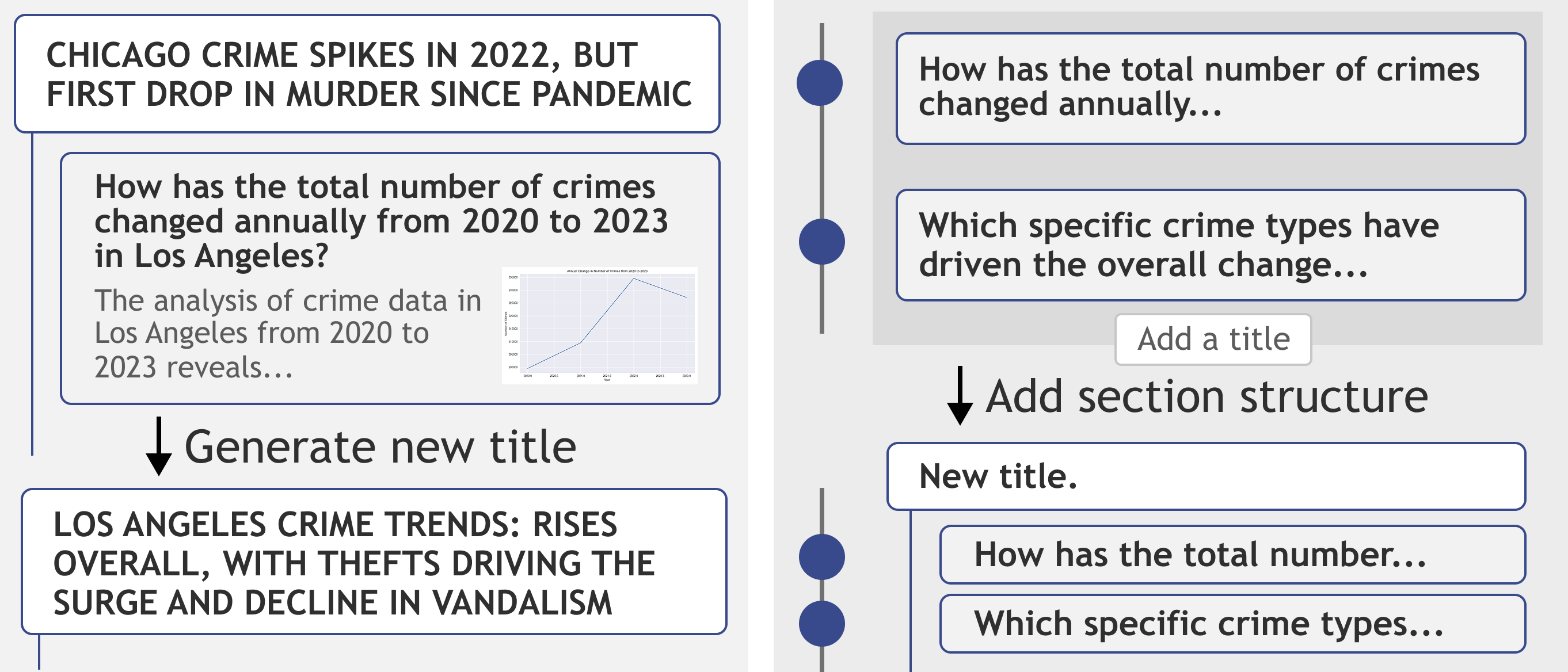}
  \caption{
    A demonstration of generating new titles and adding section structures. 
  }
  \Description{
  This figure demonstrates how users can organize the report structure in ReSpark by generating a new report title and grouping related analysis segments into sections. The left side shows the automatic generation of a new report title based on updated content. The right side illustrates how users can add section headers and reorganize segments to improve report clarity.
  }
  \label{fig:title}
\end{figure}

\textbf{Structure organization. }
After completing the data analysis, the analyst turns to structuring the report, including generating the title and organizing content into clear sections (\autoref{fig:title}).
S/he generates a report title summarizing the rise in overall crime and theft incidents, then adds and generates section headers to group the content into two coherent parts.

\begin{figure}[!htb] 
  \centering
  \includegraphics[width=\columnwidth]{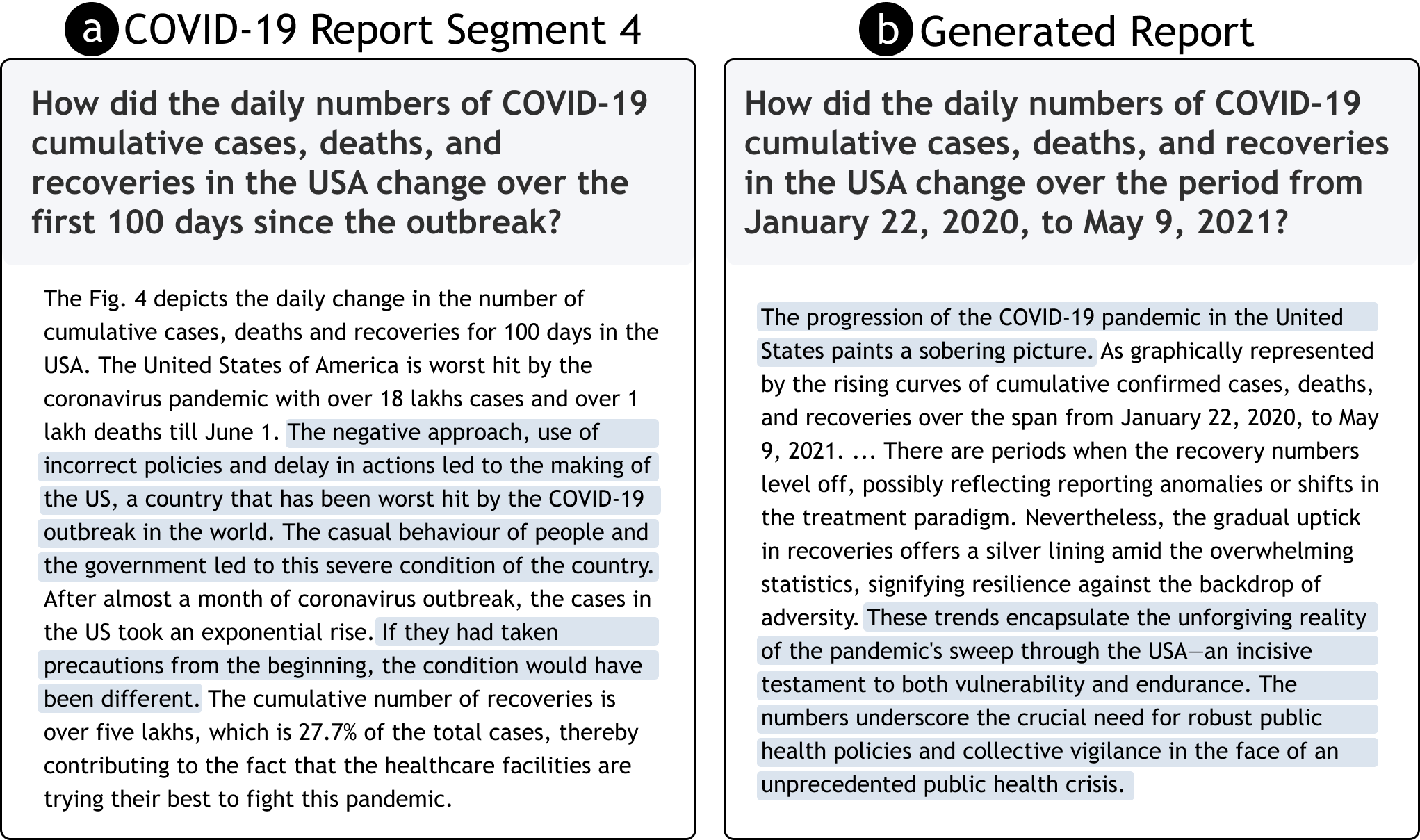}
  \caption{
    A demonstration of highlighting the sentences that serve non-data analysis purposes. 
  }
  \Description{
  This figure highlights subjective or non-data-driven statements in both a reference COVID-19 report (left) and the generated report (right). Sentences with speculative language or policy commentary are shaded to show their narrative nature. This illustrates how ReSpark can identify and flag subjective content that goes beyond data analysis.
  }
  \label{fig:highlight}
\end{figure}

\textbf{Highlight. }
For report texts, \system{} highlights sentences that serve non-data analysis purposes (\autoref{fig:highlight}). Since these sentences may lack data support, highlighting them can alert the analyst to selectively receive this information.

\section{Comparative Study}
\label{sec:comparative_study}

We conducted a comparative study to evaluate the effectiveness of the \system{} pipeline in reusing analytical logic from reference reports for new datasets. 
Our goal was to assess two aspects:
(1) whether providing a reference report improves the generated report's quality, and
(2) whether \system{}'s segmentation and step-by-step adaptation lead to better report quality than directly supplying a reference report.

To address these questions, we generated data reports under three different conditions: 
(a) using \system{} with the reference report (GPT-4o as the foundation model), 
(b) using GPT-4o directly without the reference report, and 
(c) using GPT-4o directly with the same reference report as in (a). 
This design allowed us to isolate the contributions of the reference report and the structured reuse pipeline. 
All reports were generated automatically using LLMs without human modification, ensuring a fair comparison. 
Participants were asked to review and rate each report's quality. 

\subsection{Study Setup}
\subsubsection{Participants}

We recruited 18 participants (P1–P18, 11 males and 7 females) from diverse domains, including data visualization (9), data science (4), machine learning (3), network security (1), and human-computer interaction (1).
All participants had experience reading data reports (e.g., official statistics, industry reports, or academic surveys) and self-reported their familiarity with data analysis with an average score of 3.5 on a 5-point Likert Scale. 
None of them had prior exposure to our system.

\subsubsection{Data and Reference Reports}
We selected three datasets for the study: the LA crime dataset described in \autoref{sec:usage_scenario} and two popular Kaggle datasets, a cardiovascular disease dataset and the Titanic dataset. 
These datasets were chosen for their diversity of data fields (temporal, categorical, and quantitative), high data quality, and accessibility to non-expert readers. 

For each dataset, we selected a top-ranked reference report from our repository. 
For the crime dataset, we chose the report on Chicago crime, which aligns well with both the topic and data schema. 
For the cardiovascular dataset, we selected a report on Alzheimer's disease, which is topically relevant and shares partially overlapping data fields. 
For the Titanic dataset, we used a report on voting patterns in Britain. Although it differs in topic, it shares similar data characteristics, such as demographic group analysis. 
The datasets and reference reports used in the study are listed in the supplementary materials.

\subsubsection{Report Generation}
For each dataset, we generated three versions of the report, corresponding to the three approaches described above.
\textbf{\system{}} followed the pipeline described above, which automatically segmented the reference report, adapted analytical objectives, and generated code, charts, and narratives for each segment. 
Each segment was also assigned a generated title, and a final report title was produced. 
No additional manual editing was introduced.

\textbf{GPT-4o baselines} were prompted to follow a report generation structure similar to \system{}: define objectives, generate transformations, and write insight narratives.
Specifically, we guided the model to generate report segments iteratively. In each step, it was asked to produce an analytical objective, write and execute corresponding code, generate a chart, and compose an insight narrative.
The resulting chart and text were grouped into a segment, along with a generated segment-level title.
The model was then prompted with ``Continue'' to produce the next segment. 
If errors occurred (e.g., code execution failures or incomplete output), we re-generated until valid results were obtained.
By repeating this process, the same number of segments as \system{}'s report was generated for both baselines. 
Finally, the model generated a title for the full report, consistent with the \system{} output.
The two baselines differed in their access to the reference report:

In the \textbf{GPT-only} condition, the model received only the CSV dataset and a general prompt. 
This tested its ability to generate a complete data report without any external guidance or examples, corresponding to the first goal. 

In the \textbf{GPT+reference} condition, the model additionally received the reference report's text and charts and was instructed to adapt its logic to the new data. 
Different from \system{}, the model did not receive any pre-segmented structure or extracted objectives. 
This allowed us to evaluate whether \system{}'s structured segmentation provides benefits beyond simply offering a reference example, corresponding to the second goal. 
The reports generated by the three methods and the prompts used are provided in the supplementary materials.

\subsubsection{Procedure and Tasks}
Participants were assigned to review one of three sets of data reports (each corresponding to a dataset), with each set containing three report versions for comparison.
Each set was ensured to be reviewed by six participants, totaling 18 participants across all three sets.
For their assigned set, participants were instructed to carefully read the reports and rate each one on six criteria using 5-point Likert scales: overall quality, insightfulness, logicality, chart effectiveness, text effectiveness, and chart–text consistency.
Logicality refers to the coherence and flow across segments, while chart/text effectiveness captures how well each medium conveys insights. 
Consistency measures the alignment between charts and accompanying text.
Each set required approximately 30 minutes to complete, followed by a 15-minute interview to explain their choices. 
To mitigate order effects, the order of sets and report versions within each set was counterbalanced.
The conditions were tested blindly: participants were not aware of which method generated each report.
Each participant received \$7 after the experiment. 

\begin{figure}[!htb] 
  \centering 
  \includegraphics[width=\columnwidth]{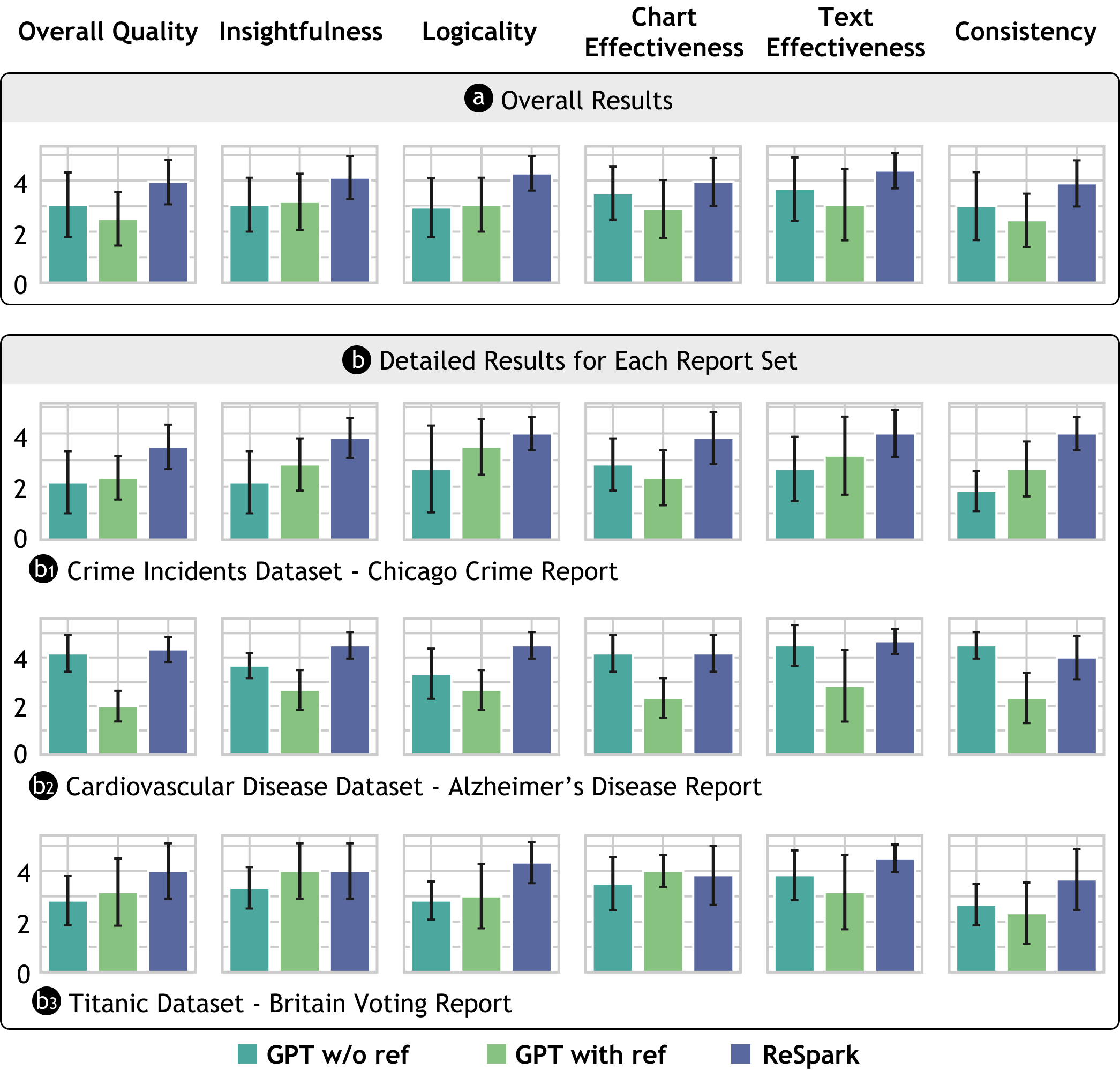}
  \caption{
  Results of the 5-point Likert scale questionnaire in the comparative study. (a) Overall results summarizing all ratings across the three report sets. (b) Detailed results for each report set. Error bars indicate standard deviations (SD).
  }
  \Description{
  A grid of bar charts displays Likert scale ratings from the comparative study evaluating report quality. Subfigure (a) summarizes all user ratings across three report sets. Subfigures (b1), (b2), and (b3) present results for each report set individually. Each chart includes bars representing three versions of reports: baseline (teal), ablation (light green), and ReSpark (dark blue), with error bars showing standard deviations. ReSpark consistently receives higher ratings across most criteria.
  }
  \label{fig:evaluation} 
\end{figure}

\subsection{Results and Discussions}

Overall, participants gave higher average ratings to the reports generated by \system{} across all aspects (\autoref{fig:evaluation}a). 
We used a Friedman Test to examine whether there were significant differences across the three methods.
Significant differences were observed in five aspects: overall quality, logicality, chart effectiveness, text effectiveness, and consistency ($\chi^2$ = 8.71, 8.45, 7.66, 6.82, and 7.24, respectively; $p < .05$). 
As post-hoc analyses, we conducted two-sided Wilcoxon signed-rank tests on paired responses between each pair of methods (\system{}, GPT-only, and GPT+reference), with Bonferroni correction~\cite{Bonferroni} for multiple comparisons.
Significant differences were found between \system{} and both baselines in logicality ($p < .05$), and between \system{} and the GPT+reference baseline in overall quality, text effectiveness, and consistency ($p < .05$).
We further summarized the important feedback as follows. 

\subsubsection{Progressive Logical Flow}
\system{} reports received significantly higher ratings for logicality compared to GPT-only and GPT+reference baselines.
This was also the most frequently mentioned strength across interviews, with 13 out of 18 participants highlighting the logical progression of \system{}'s reports compared to the baseline versions across all three sets. 
Specifically, 9 participants remarked that \system{} reports exhibited a ``clear overall-to-specific structure, making the logic of the report more apparent,'' and 8 noted that paragraphs ``progress step by step, with a logical and cohesive flow.'' 
P8 pointed out how the report on the cardiovascular disease dataset followed such a structure: ``It starts with the prevalence of the disease, then moves to find the leading factors, and then analyzes each key factor in detail.'' 
This perceived logicality also contributed to higher ratings for overall quality and insightfulness, with 8 participants linking deeper analysis to stronger insights and quality.

In contrast, both baselines received feedback indicating that their narrative structure was ``flat'' and ``lacked a stepwise, progressive flow.''
13 participants noted that the paragraphs ``had few connections'' and felt ``disjointed'' while reading them. 
P2 and P14 remarked that baseline reports ``tend to analyze more dimensions, but each factor was not examined in depth,'' reflecting a shallow, breadth-first style. 

\subsubsection{Impact of Reference Report and Structured Reuse Pipeline} 
There was no consistent preference between the two GPT-4o baselines among all six criteria, and no statistically significant differences were found between them in any aspect.
Scores for baselines with and without the reference report varied across datasets and dimensions, suggesting that simply providing a reference report for the model does not lead to clearly better outcomes. 

Upon closer inspection, we found that the baseline provided with the reference report did not clearly inherit the reference report's logical structure. 
As noted earlier, participants found both baseline reports lacking in logical coherence. 
Since the key difference between GPT with reference and \system{} lies in the pre-processed segmentation and step-by-step adaptation, these findings suggest that this structured reuse pipeline helps retain and transfer the logical structure of the reference report, resulting in more coherent analytical flow.

\subsubsection{Issues in Chart Designs and Text-Chart Consistency} 
\begin{figure}[!htb] 
  \centering
  \includegraphics[width=\columnwidth]{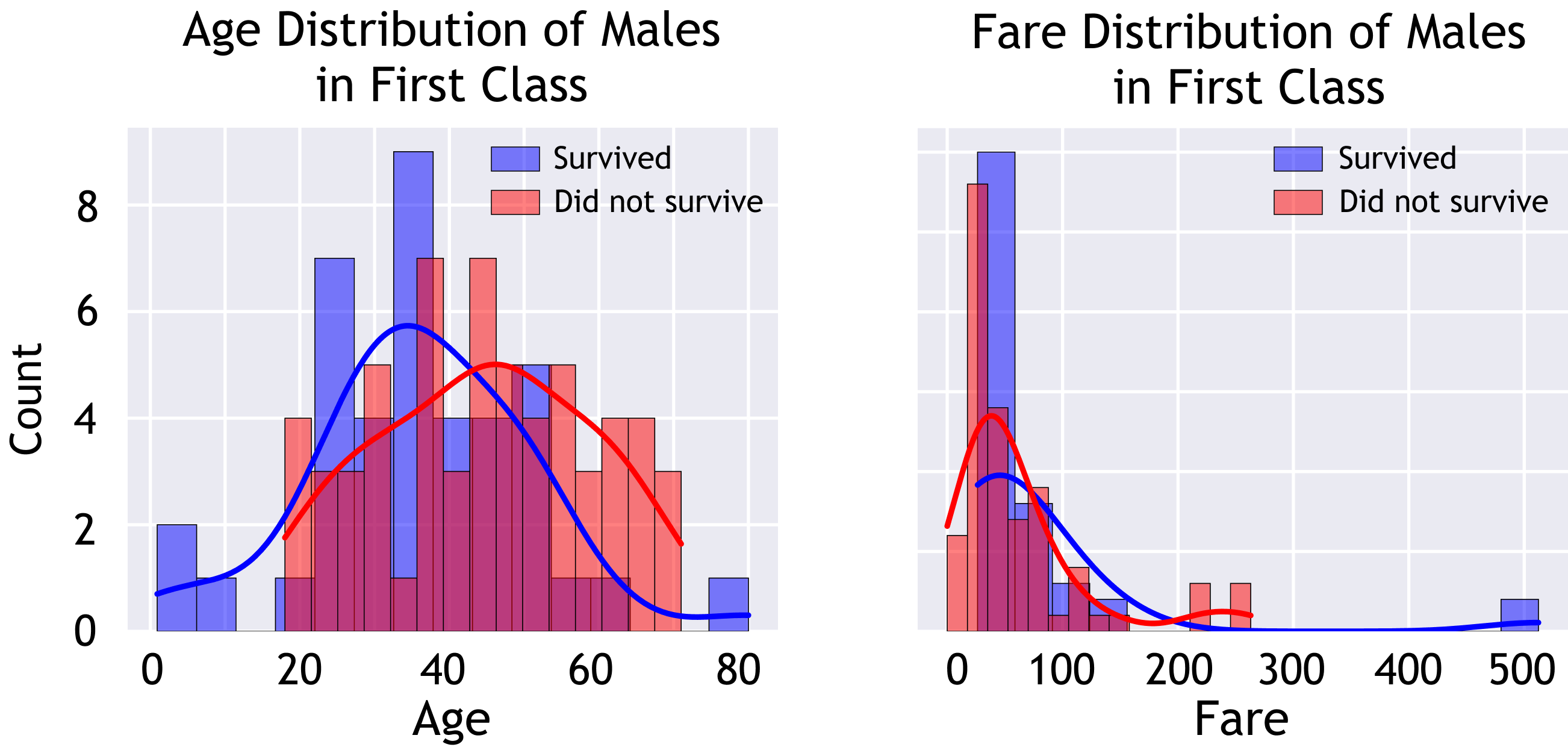}
  \caption{
    Example of an inappropriate chart generated in the Titanic dataset report, where the distinction between survivors and non-survivors is made using two colors. 
  }
  \Description{
  An example of an inappropriate chart from the Titanic dataset report. Two histograms are overlaid using color to distinguish survivors (blue) and non-survivors (red). The color scheme alone makes it difficult to interpret overlapping data, leading to potential misinterpretation. This highlights the importance of effective visual encoding in automatically generated charts.
  }
  \label{fig:compararive_error_example.png}
\end{figure}

While \system{} received the highest average ratings overall, it showed lower average scores in specific aspects of certain report sets, including chart–text consistency in the Cardiovascular Disease dataset (\autoref{fig:evaluation}b1) and chart effectiveness in the Titanic dataset (\autoref{fig:evaluation}b3).
These ratings stem from two types of issues with distinct causes: 
(1) Chart–text contradiction (1 case). 
In the cardiovascular disease report, the text stated that a certain age group had the highest cholesterol level, while the chart showed it was the second highest. 
This mismatch resulted from incorrect chart interpretation during text generation.
(2) Unclear visual design (2 cases). 
In the Titanic report, although participants appreciated the depth of analysis, one chart (\autoref{fig:compararive_error_example.png}) was considered overly complex. 
P15 suggested that it would be clearer if survival rates were plotted on the y-axis rather than using two colors to differentiate survivals. 
Another case is about misleading visual elements: a tick label of the year with a decimal number 2020.5 in the crime report. 
Such unclear visual designs were caused by inappropriate styling parameters during code generation.

These issues suggest that \system{} does not fully eliminate the potential errors that could arise from GPT-generated content. 
To help users identify and fix such issues, \system{} shows transformed data, charts, and text to support checking each component, and users can customize the charts and text to correct them.
However, these issues reflect broader challenges such as consistency checking and LLM hallucination. 
We further discuss this limitation in the discussion section.

\subsubsection{Depth vs. Breadth Trade-off}
Some participants further pointed out the trade-off between the depth and breadth of analysis. 
While \system{} was praised for in-depth and logical exploration, P1 noted its narrower focus: the crime report, for instance, ``only focused on a limited set of data fields.'' 
This reflects a structural trade-off: deeper analysis within a fixed report length may lead to a narrower focus. 
It's akin to a tree with the same number of nodes: a deeper tree has a smaller breadth. 
Since \system{} inherits the logical structure from the reference report, it may naturally generate overly narrow reports when the reference focuses on only a few data fields, and the target dataset contains many more.
This underscores the importance of \system{}'s interactive interface, which enables users to identify un-analyzed fields and flexibly add or remove segments.
\section{Usability Study}
\label{sec:user_study}

We conducted a user study to evaluate the usability of \system{}. 
Participants were required to use \system{} to generate a complete data report for a dataset by selecting a reference report of their choice and adapting it accordingly.
They were also asked to provide feedback on their experience with the system.
In this study, \system{} leveraged GPT-4o as the foundation model.

\subsection{Study Setup}

\subsubsection{Participants}
We recruited 12 participants (U1-U12, 9 males and 3 females) from various backgrounds, including data science (5), computer science (3), mathematics (2), sports science (1), and biosystems engineering (1). 
Most participants were familiar with data analysis, with an average self-reported score of 3.25 on a 5-point Likert Scale. 
All of them had experience reading data reports.

\subsubsection{Tasks}
Each participant was tasked with using \system{} to generate a complete data report based on an assigned dataset, following two main requirements:

\textbf{Freely select a reference report.}
Each participant was asked to select a reference report freely from a set of 8 candidates, using the system's ranking, summarized information, and preview features as guidance.
Participants were encouraged to carefully read and compare the available reports and were free to switch to another option if they were unsatisfied with their choice.

\textbf{Generate a complete report.}
Using the selected reference, participants were asked to generate a complete report with \system{}, including producing segments and organizing the report structure. 
They could make any modifications during this process, including regenerating model responses, manually editing analytical objectives or text, and inserting or removing segments.
The final report was required to contain at least four segments.

\subsubsection{Datasets and Candidate Reference Reports}
We used the same three datasets as in the comparative study (\autoref{sec:comparative_study}): the LA crime dataset, the CDC disease dataset, and the Titanic dataset.
Participants were randomly assigned one of three datasets, ensuring balanced coverage across the study.

To support reference selection, participants were given 8 candidate reports from diverse domains (e.g., health, education, security).
We limited the set to 8 reports to ensure participants could evaluate each option.
This setup allowed us to evaluate the report ranking mechanism and understand participants' selection strategies. 

Moreover, these datasets and reports were selected to cover a range of alignment scenarios:
The \textbf{LA crime dataset} is closely aligned with a Chicago crime report in both topic and data fields.
The \textbf{CDC disease dataset} is topically relevant to a report on Alzheimer's disease but has lower field similarity (e.g., the Alzheimer's report includes geo-location, while CDC data does not).
The \textbf{Titanic dataset} is not topically aligned with any candidate report but shares demographic fields (e.g., age, gender) with some reports, such as a report on election voting.
This diversity allowed us to test \system{}'s performance across varying levels of field similarity and topic relevance.
The datasets and reference reports used in the study are listed in the supplementary materials.

\subsubsection{Procedure}
The entire experiment lasted approximately 70 minutes. 
It began with a 3-minute introduction to the concept of reference-based report generation, followed by a 10-minute tutorial on \system{} interactions. 
Participants then explored the interface freely for 3 minutes.

After that, participants were tasked with using \system{} to generate a data report, which took approximately 35 minutes. 
All interactions were logged during this process. 
After completing their report, participants filled out a post-study questionnaire that included the System Usability Scale (SUS)~\cite{SUS} and 5-point Likert ratings for specific system features.
Finally, they participated in a 15-minute semi-structured interview to share qualitative feedback, such as the factors they considered when choosing the reference report.
Each participant received a \$10 compensation for their time.

\begin{figure}[!htb] 
  \centering
  \includegraphics[width=\columnwidth]{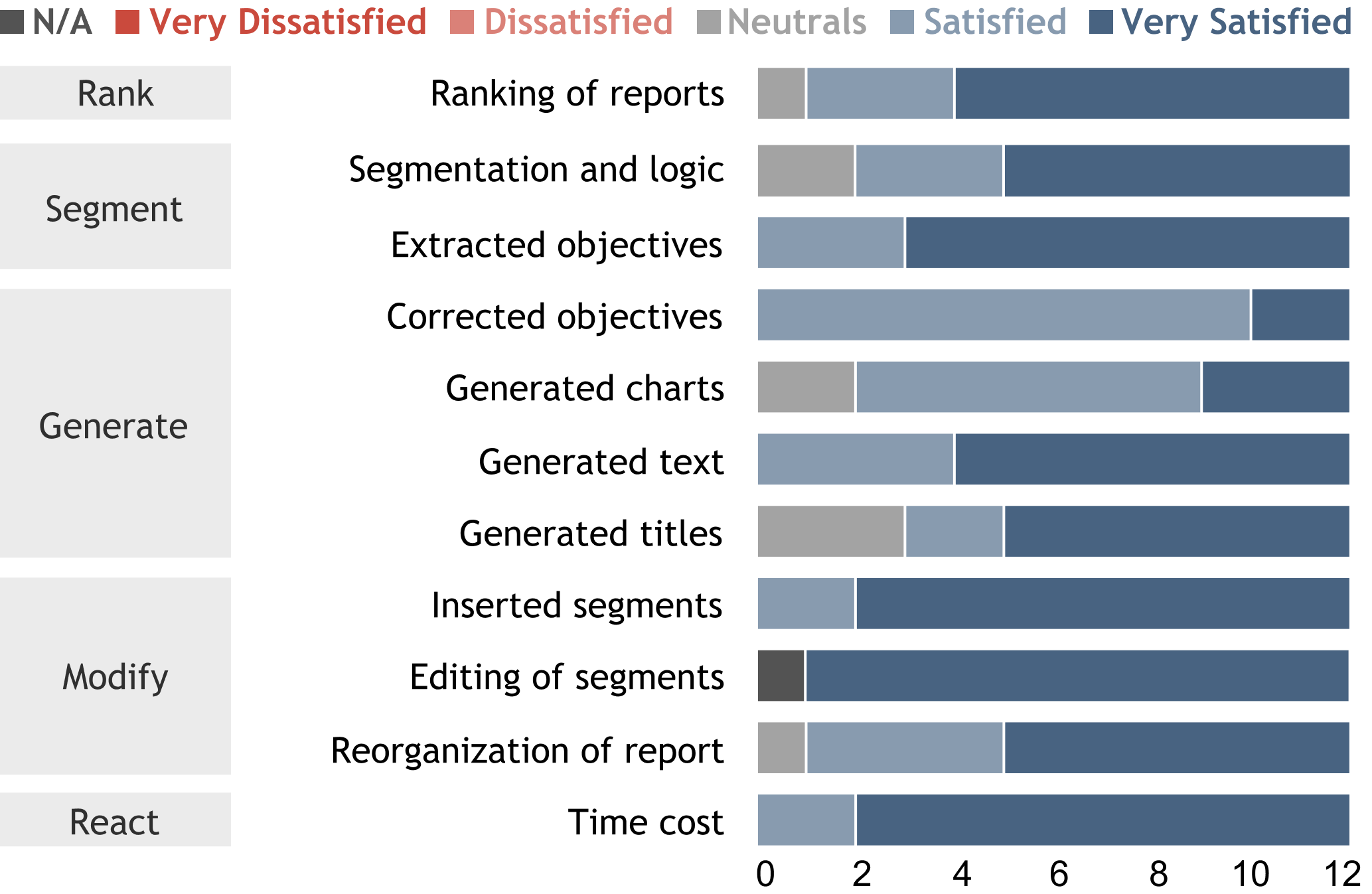}
  \caption{User ratings for different features of \system{}.}
  \Description{
  User satisfaction ratings for different features of ReSpark. Each bar represents responses on a 5-point Likert scale, from ``Very Dissatisfied'' to ``Very Satisfied.'' Across features such as ranking, segmentation, generation, modification, and time cost, the majority of users reported being satisfied or very satisfied.
  }
  \label{fig:user_study_result}
\end{figure}

\begin{figure*}[!htb] 
  \centering
  \includegraphics[width=\textwidth]{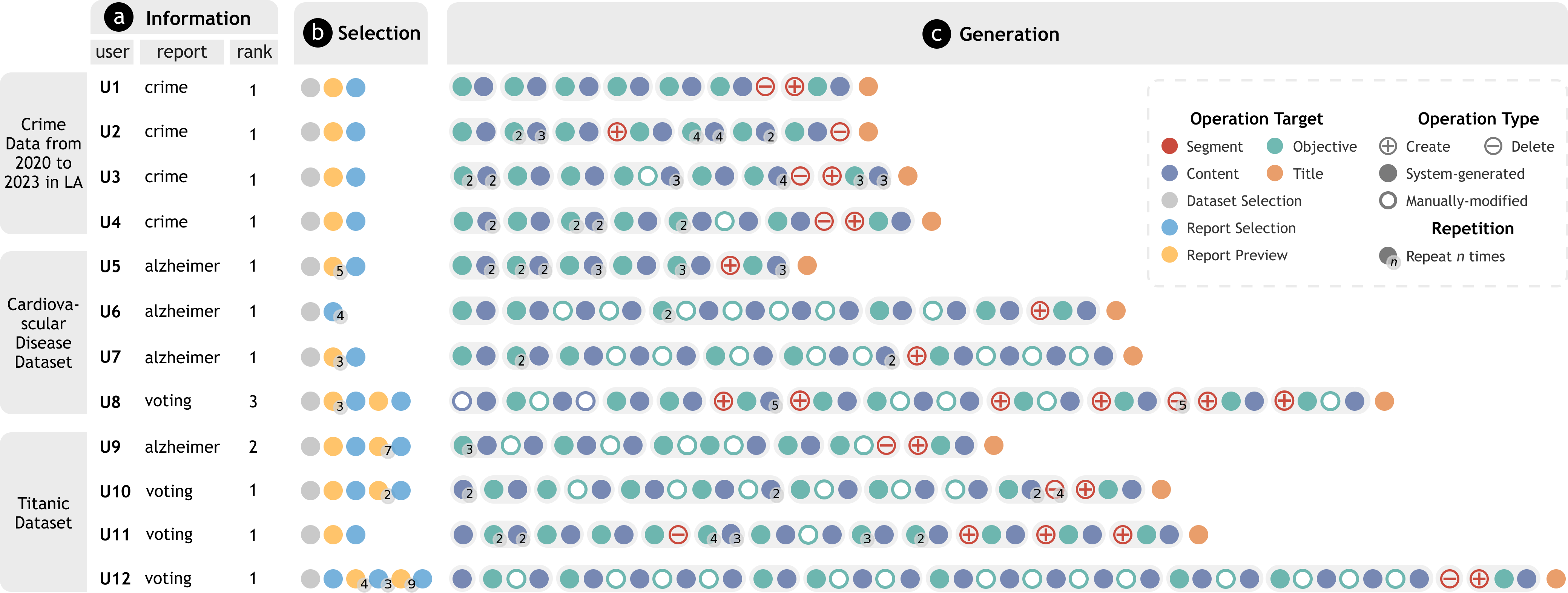}
  \caption{Interaction sequences from the usability study, including user information with assigned datasets and selected reference reports with their ranks (a) and interaction logs during reference report selection (b) and new report generation (c).}
  \Description{Interaction log from the usability study. Panel (a) shows user assignments, selected reference reports, and their ranks. Panel (b) visualizes participants' interactions during reference report selection. Panel (c) presents the full interaction sequences during report generation, including system-generated and user-modified objectives, content, and segment edits.}
  \label{fig:user_involvement}
\end{figure*}

\subsection{Results}

The user ratings on system features are illustrated in~\autoref{fig:user_study_result}, and user interaction sequences are visualized in~\autoref{fig:user_involvement}. We summarize the key findings from both quantitative results and qualitative feedback below.

\subsubsection{Ease of Use}
\system{} received an overall SUS score of 88.96, with scores across the three datasets as follows: crime (86.88), disease (91.88), and Titanic (88.13). 
This suggests consistently high usability, as SUS scores above 80.3 are generally regarded as being in the top 10\%~\cite{SUS}. 

Most participants agree that \system{} is easy to use. 
Six participants described it as ``intuitive'' and ``having a low learning curve.''
Additionally, all participants agreed that the system response time was acceptable. 
U2, U3, and U9 specifically noted that the real-time feedback made the generation time feel manageable.
U9 remarked, ``Seeing the results being generated gradually reassures me that \system{} is progressing. This allows me to stay informed about the model's generation process.''

\subsubsection{Reference Report Ranking and Selection}
All participants agree that the \system{}'s ranking of the alternative reports is suitable. 
As shown in~\autoref{fig:user_involvement}a, most participants selected the first-ranked report, and all chose the top three, suggesting the effectiveness of our ranking mechanism.
We further summarize their interactive behaviors and considerations for report selection as follows. 

\textbf{Ranking and summary information guide selection, preview completes the decision. } 
During report selection, participants were initially provided with a report list displaying the title, topic, and predicted data fields for each candidate report, along with the option to preview full content (\autoref{fig:retrieve}).
As shown in~\autoref{fig:user_involvement}b, most participants did not preview and select all eight reports. 
Instead, they used the summary list to form initial impressions and selectively previewed a few candidates that appeared suitable.
This behavior suggests that the summarized information and ranking effectively helped participants narrow down their options.
Previewing remained essential for final decisions, as most participants used it before making their choice.
This behavior was more common with the disease and Titanic datasets, possibly due to their lower alignment with the available reports.

\textbf{Analysis method applicability guide reference selection.}
All participants considered both field similarity and topic relevance as important criteria, but 9 out of 12 prioritized field similarity.
They also found the predicted data fields in the report list (\autoref{fig:retrieve}) more useful than topic labels.
This preference stemmed from a deeper concern: whether the report offered high-level analysis tasks—such as trend analysis or group comparison—that could be applied to the target dataset.
As U10 and U11 noted, ``similar and overlapping data fields are likely to use similar analytical tasks,'' making field similarity a practical signal. 
In contrast, topic labels were seen as ``providing general context '', but ``not necessarily imply transferable analytical tasks.''

While many viewed field similarity as a useful proxy for analysis tasks, some participants noted its limitations.
As U11 noted, ``Even when data fields differ, the same method may still apply. For example, the Titanic dataset has a unique `cabin' variable not found in other reports, but it still fits group comparison analysis.''
One possible approach is to mine and summarize the high-level analytical methods used in each report and assess their applicability to the target dataset, which requires additional domain knowledge. 

\subsubsection{User Involvement}
While most participants agreed that the generated objectives and content were generally correct and useful (\autoref{fig:user_study_result}), all participants used various interactions to customize their reports (\autoref{fig:user_involvement}).
We summarize their behaviors and underlying motivations as follows.

\textbf{Modifying objectives for detailed intent.}
Most participants frequently adjusted the objectives.
However, participants indicated that these modifications were not driven by errors but to express more specific analytical goals. 
As U7 commented, ``At first, I didn't know exactly what to analyze, but after seeing the generated objectives, it inspired me to form clearer goals. ''

Edits of these objectives commonly involved changing data fields or specifying chart types.
U12, who used this feature most extensively, appreciated it: ``I just need to make small modifications, like changing a data field or adding a requirement for the chart type.''
U8 even incorporated statistical tests. 
This feedback demonstrates that the flexibility of modifying objectives helps users face precise analytical needs. 

\textbf{Inserting or deleting segments to support deeper coverage.}
All participants used the feature of inserting or removing report segments, typically in later stages of their generation.
Eight described it as ``very necessary,'' especially when combined with the interface showing how often data fields had been used.
U6 noted, ``These statistics help me identify un-analyzed data fields, so I can manually insert segments for them.''

U11 further described a gradual process, ``The reference report gives me a warm start. As I continue, my analysis intent becomes clearer, so I modify the objectives more often. Eventually, I add segments manually for more complete and deeper analysis.''
This pattern aligns with the interaction logs: early edits were minimal, followed by more substantial and frequent customization as users' goals became clearer.

\textbf{Using regeneration because of uncertainty or convenience.}
Some participants used regeneration more frequently than direct edits.
Two main reasons emerged: U5 mentioned that he was often unsure about his desired results, but could decide whether a result worked once he saw it. 
As a result, he preferred to regenerate outputs until the result felt appropriate. 
On the other hand, U11 preferred not to type and would regenerate until a satisfying result appeared. 
He suggested the system offer more structured controls, like interactive options for chart types, to reduce reliance on manual input.
We plan to incorporate this feature in future iterations.
\section{Discussions}
In this section, we reflect on the broader implications of our work and outline its limitations and directions for future research.

\subsection{Implications}

We discuss the implications of our work from two aspects. 

\textbf{Reuse as a design paradigm for data storytelling.}
While prior work has explored the reuse of glyphs~\cite{ying2022metaglyph} and code~\cite{epperson2022strategiesReuseTeam} for new data or scenarios, applying this paradigm to data storytelling remains underexplored. 
ReSpark demonstrates how existing reports can be reused to capture embedded human expertise and guide new tasks.
Given the abundance of human-authored materials in data storytelling, this paradigm offers potential for broader application.

\textbf{Leveraging underlying knowledge beyond surface text.}
ReSpark enables LLMs to extract high-level analytical logic from reference reports, going beyond surface-level retrieval as in RAG approaches. 
Future work could explore how to extract and adapt deeper underlying knowledge in existing materials, such as analytic intent, reasoning patterns, and methodological framing, to scaffold complex, multi-step tasks in domains like data analysis.

\subsection{Limitations and Future Work}

This subsection discusses the current limitations of the \system{} framework and outlines opportunities for future improvements.

\textbf{Handling multiple data sources and references.}
\system{} currently operates by taking a single dataset and a reference report as input. 
However, in practice, creating more complex data journalism articles (such as investigations and in-depth inverstigations~\cite{designpatterns2025hao}) often requires synthesizing insights from diverse datasets and consulting multiple reference materials~\cite{li2023ai, cao2023dataparticles}. 
These scenarios may also involve highly specialized domains where LLMs lack sufficient domain knowledge or contextual understanding.

Supporting such cases would require new mechanisms to: (1) extract and merge analytical objectives from multiple references, (2) identify relevant fields across datasets to jointly support these objectives, and (3) incorporate external domain knowledge, evidence, or expert insights to enable deeper analysis. 
Future work could develop pipelines that integrate analysis logic across diverse sources, incorporate domain-specific knowledge, and synthesize coherent, insight-rich reports from these heterogeneous inputs.

\textbf{Extend to interactive presentations of analysis.}
\system{} currently produces static outputs, including plain text and charts. 
One of the next phases of development is to support interactive outputs. 
Prior work has explored chart–text linking~\cite{sultanum2023datatales, chen2022crossdata} and animated transitions across narrative segments~\cite{cao2023dataparticles, morth2022scrollyvis}, which can be integrated for more impressive data reports. 

Beyond these forms, more powerful formats, such as multi-view dashboards~\cite{dengDashBotInsightDrivenDashboard2022} and visual analysis interfaces of multiple coordinated views (MCVs)~\cite{zhaoLightVALightweightVisual2024}, offer greater interactivity. 
Future work could explore generating such interfaces by referencing existing visual analytics systems. 
This would require extracting analytical logic from interface components and user interaction workflows instead of linear narrative-based structures.

\textbf{Support for data pre-processing.} 
\system{} is designed to analyze data directly, progressing seamlessly from one analytical objective to another. 
Therefore, the input data needs to be clean and pre-processed, which does not need additional data wrangling and is ready for analysis. 
However, in real-world scenarios, data often requires pre-processing, such as handling missing values, correcting inconsistencies, and normalization, before analysis can begin.
While LLMs are capable of generating data wrangling code~\cite{huang2023nl2rigel}, reliably detecting and addressing data quality issues remains a complex task, which we consider to be out of the current scope of \system{}. 
Future work could integrate an additional module to assist with data cleaning and wrangling before beginning analysis. 
This module could leverage existing techniques for automated data cleaning and wrangling~\cite{luo2024ferry}, improving the system's ability to handle raw or semi-structured datasets and expanding its applicability across a broader range of real-world scenarios.

\textbf{Smart verification and correction of LLM-generated results.}
Since all content in \system{}, including objectives, code, and text, is generated by LLMs, verifying its correctness is important. 
Our comparative study revealed issues like mismatches between charts and text, and unclear visual designs.
To address this, \system{} supports lightweight review by showing intermediate code, transformed data, and charts, and allowing users to adjust them manually. 
However, identifying these issues remains not easy, as it requires mentally aligning the chart, data, and text, which imposes cognitive overhead. 
Future work could link the text, data, and chart elements~\cite{chen2022crossdata, sultanum2023datatales} to surface potential inconsistencies. 
Automated evaluation frameworks like VisEval~\cite{chen2024viseval} can also help flag chart design issues across aspects such as validity, readability, and legality.
Building on such detections, LLMs could be prompted to regenerate the results, enabling iterative refinement with less manual effort.

\section{Conclusion}

We present \system{}, an LLM-driven method that assists users in generating data reports by reusing the analytical logic of a selected reference report. 
By decomposing the reference report into a sequence of interconnected segments with inferred analytical objectives and dependencies, \system{} adapts this analysis logic to new datasets through step-by-step generation.
An interactive interface is designed to allow users to inspect, modify, and extend the generated report as needed. 
The effectiveness of \system{} is evaluated through comparative and user studies. 

\begin{acks}
This work was supported by Zhejiang Provincial Natural Science Foundation of China under Grant No. LD25F020003 and NSFC (U22A2032, 62402428), and partially supported by ZJU Kunpeng \& Ascend Center of Excellence. The author also gratefully acknowledges the support of Zhejiang University Education Foundation Qizhen Scholar Foundation.
\end{acks}

\bibliographystyle{ACM-Reference-Format}
\bibliography{ref}


\begin{thebibliography}{67}


\ifx \showCODEN    \undefined \def \showCODEN     #1{\unskip}     \fi
\ifx \showISBNx    \undefined \def \showISBNx     #1{\unskip}     \fi
\ifx \showISBNxiii \undefined \def \showISBNxiii  #1{\unskip}     \fi
\ifx \showISSN     \undefined \def \showISSN      #1{\unskip}     \fi
\ifx \showLCCN     \undefined \def \showLCCN      #1{\unskip}     \fi
\ifx \shownote     \undefined \def \shownote      #1{#1}          \fi
\ifx \showarticletitle \undefined \def \showarticletitle #1{#1}   \fi
\ifx \showURL      \undefined \def \showURL       {\relax}        \fi
\providecommand\bibfield[2]{#2}
\providecommand\bibinfo[2]{#2}
\providecommand\natexlab[1]{#1}
\providecommand\showeprint[2][]{arXiv:#2}

\bibitem[Bar~El et~al\mbox{.}(2020)]%
        {bar2020automatically}
\bibfield{author}{\bibinfo{person}{Ori Bar~El}, \bibinfo{person}{Tova Milo}, {and} \bibinfo{person}{Amit Somech}.} \bibinfo{year}{2020}\natexlab{}.
\newblock \showarticletitle{Automatically generating data exploration sessions using deep reinforcement learning}. In \bibinfo{booktitle}{\emph{Proceedings of the ACM SIGMOD International Conference on Management of Data}}. \bibinfo{publisher}{Association for Computing Machinery}, \bibinfo{address}{New York, NY, USA}, \bibinfo{pages}{1527--1537}.
\newblock
\href{https://doi.org/10.1145/3318464.3389779}{doi:\nolinkurl{10.1145/3318464.3389779}}


\bibitem[Batch and Elmqvist(2017)]%
        {batch2017interactive}
\bibfield{author}{\bibinfo{person}{Andrea Batch} {and} \bibinfo{person}{Niklas Elmqvist}.} \bibinfo{year}{2017}\natexlab{}.
\newblock \showarticletitle{The interactive visualization gap in initial exploratory data analysis}.
\newblock \bibinfo{journal}{\emph{IEEE Transactions on Visualization and Computer Graphics}} \bibinfo{volume}{24}, \bibinfo{number}{1} (\bibinfo{year}{2017}), \bibinfo{pages}{278--287}.
\newblock
\href{https://doi.org/10.1109/TVCG.2017.2743990}{doi:\nolinkurl{10.1109/TVCG.2017.2743990}}


\bibitem[Battle and Heer(2019)]%
        {battle2019characterizing}
\bibfield{author}{\bibinfo{person}{Leilani Battle} {and} \bibinfo{person}{Jeffrey Heer}.} \bibinfo{year}{2019}\natexlab{}.
\newblock \showarticletitle{Characterizing exploratory visual analysis: A literature review and evaluation of analytic provenance in tableau}.
\newblock \bibinfo{journal}{\emph{Computer Graphics Forum}} \bibinfo{volume}{38}, \bibinfo{number}{3} (\bibinfo{year}{2019}), \bibinfo{pages}{145--159}.
\newblock
\href{https://doi.org/10.1111/cgf.13678}{doi:\nolinkurl{10.1111/cgf.13678}}


\bibitem[Behrens(1997)]%
        {behrens1997principles}
\bibfield{author}{\bibinfo{person}{John~T Behrens}.} \bibinfo{year}{1997}\natexlab{}.
\newblock \showarticletitle{Principles and procedures of exploratory data analysis}.
\newblock \bibinfo{journal}{\emph{Psychological Methods}} \bibinfo{volume}{2}, \bibinfo{number}{2} (\bibinfo{year}{1997}), \bibinfo{pages}{131}.
\newblock
\href{https://doi.org/10.1037/1082-989X.2.2.131}{doi:\nolinkurl{10.1037/1082-989X.2.2.131}}


\bibitem[Benavoli et~al\mbox{.}(2016)]%
        {Bonferroni}
\bibfield{author}{\bibinfo{person}{Alessio Benavoli}, \bibinfo{person}{Giorgio Corani}, {and} \bibinfo{person}{Francesca Mangili}.} \bibinfo{year}{2016}\natexlab{}.
\newblock \showarticletitle{Should we really use post-hoc tests based on mean-ranks?}
\newblock \bibinfo{journal}{\emph{J. Mach. Learn. Res.}} \bibinfo{volume}{17}, \bibinfo{number}{1} (\bibinfo{date}{Jan.} \bibinfo{year}{2016}), \bibinfo{pages}{152–161}.
\newblock
\showISSN{1532-4435}


\bibitem[Brooke(1996)]%
        {SUS}
\bibfield{author}{\bibinfo{person}{John Brooke}.} \bibinfo{year}{1996}\natexlab{}.
\newblock \bibinfo{booktitle}{\emph{{SUS: A Quick and Dirty Usability Scale}}}.
\newblock \bibinfo{pages}{189--194}.
\newblock


\bibitem[Cao et~al\mbox{.}(2023)]%
        {cao2023dataparticles}
\bibfield{author}{\bibinfo{person}{Yining Cao}, \bibinfo{person}{Jane~L E}, \bibinfo{person}{Zhutian Chen}, {and} \bibinfo{person}{Haijun Xia}.} \bibinfo{year}{2023}\natexlab{}.
\newblock \showarticletitle{{DataParticles}: Block-based and language-oriented authoring of animated unit visualizations}. In \bibinfo{booktitle}{\emph{Proceedings of the CHI Conference on Human Factors in Computing Systems}}. \bibinfo{publisher}{Association for Computing Machinery}, \bibinfo{address}{New York, NY, USA}, \bibinfo{pages}{1--15}.
\newblock
\href{https://doi.org/10.1145/3544548.3581472}{doi:\nolinkurl{10.1145/3544548.3581472}}


\bibitem[Chen et~al\mbox{.}(2025)]%
        {chen2024viseval}
\bibfield{author}{\bibinfo{person}{Nan Chen}, \bibinfo{person}{Yuge Zhang}, \bibinfo{person}{Jiahang Xu}, \bibinfo{person}{Kan Ren}, {and} \bibinfo{person}{Yuqing Yang}.} \bibinfo{year}{2025}\natexlab{}.
\newblock \showarticletitle{VisEval: A Benchmark for Data Visualization in the Era of Large Language Models}.
\newblock \bibinfo{journal}{\emph{IEEE Transactions on Visualization and Computer Graphics}} \bibinfo{volume}{31}, \bibinfo{number}{1} (\bibinfo{year}{2025}), \bibinfo{pages}{1301--1311}.
\newblock
\href{https://doi.org/10.1109/TVCG.2024.3456320}{doi:\nolinkurl{10.1109/TVCG.2024.3456320}}


\bibitem[Chen et~al\mbox{.}(2023)]%
        {chen2023calliopeNet}
\bibfield{author}{\bibinfo{person}{Qing Chen}, \bibinfo{person}{Nan Chen}, \bibinfo{person}{Wei Shuai}, \bibinfo{person}{Guande Wu}, \bibinfo{person}{Zhe Xu}, \bibinfo{person}{Hanghang Tong}, {and} \bibinfo{person}{Nan Cao}.} \bibinfo{year}{2023}\natexlab{}.
\newblock \showarticletitle{{Calliope-Net}: Automatic generation of graph data facts via annotated node-link diagrams}.
\newblock \bibinfo{journal}{\emph{IEEE Transactions on Visualization and Computer Graphics}} \bibinfo{volume}{30}, \bibinfo{number}{1} (\bibinfo{year}{2023}), \bibinfo{pages}{562--572}.
\newblock
\href{https://doi.org/10.1109/TVCG.2023.3326925}{doi:\nolinkurl{10.1109/TVCG.2023.3326925}}


\bibitem[Chen and Xia(2022)]%
        {chen2022crossdata}
\bibfield{author}{\bibinfo{person}{Zhutian Chen} {and} \bibinfo{person}{Haijun Xia}.} \bibinfo{year}{2022}\natexlab{}.
\newblock \showarticletitle{{CrossData}: Leveraging text-data connections for authoring data documents}. In \bibinfo{booktitle}{\emph{Proceedings of the CHI Conference on Human Factors in Computing Systems}}. \bibinfo{publisher}{Association for Computing Machinery}, \bibinfo{address}{New York, NY, USA}, \bibinfo{pages}{1--15}.
\newblock
\href{https://doi.org/10.1145/3491102.3517485}{doi:\nolinkurl{10.1145/3491102.3517485}}


\bibitem[Cheng et~al\mbox{.}(2025)]%
        {cheng2024snil}
\bibfield{author}{\bibinfo{person}{Liqi Cheng}, \bibinfo{person}{Dazhen Deng}, \bibinfo{person}{Xiao Xie}, \bibinfo{person}{Rihong Qiu}, \bibinfo{person}{Mingliang Xu}, {and} \bibinfo{person}{Yingcai Wu}.} \bibinfo{year}{2025}\natexlab{}.
\newblock \showarticletitle{SNIL: Generating Sports News From Insights With Large Language Models}.
\newblock \bibinfo{journal}{\emph{IEEE Transactions on Visualization and Computer Graphics}} \bibinfo{volume}{31}, \bibinfo{number}{7} (\bibinfo{year}{2025}), \bibinfo{pages}{3973--3986}.
\newblock
\href{https://doi.org/10.1109/TVCG.2024.3392683}{doi:\nolinkurl{10.1109/TVCG.2024.3392683}}


\bibitem[Cheng et~al\mbox{.}(2023)]%
        {cheng2023gpt4analyst}
\bibfield{author}{\bibinfo{person}{Liying Cheng}, \bibinfo{person}{Xingxuan Li}, {and} \bibinfo{person}{Lidong Bing}.} \bibinfo{year}{2023}\natexlab{}.
\newblock \showarticletitle{Is {GPT}-4 a Good Data Analyst?}. In \bibinfo{booktitle}{\emph{Findings of the Association for Computational Linguistics: EMNLP 2023}}. \bibinfo{publisher}{Association for Computational Linguistics}, \bibinfo{address}{Singapore}, \bibinfo{pages}{9496--9514}.
\newblock
\href{https://doi.org/10.18653/v1/2023.findings-emnlp.637}{doi:\nolinkurl{10.18653/v1/2023.findings-emnlp.637}}


\bibitem[Deng et~al\mbox{.}(2022)]%
        {dengDashBotInsightDrivenDashboard2022}
\bibfield{author}{\bibinfo{person}{Dazhen Deng}, \bibinfo{person}{Aoyu Wu}, \bibinfo{person}{Huamin Qu}, {and} \bibinfo{person}{Yingcai Wu}.} \bibinfo{year}{2022}\natexlab{}.
\newblock \showarticletitle{DashBot: Insight-Driven Dashboard Generation Based on Deep Reinforcement Learning}.
\newblock \bibinfo{journal}{\emph{IEEE Transactions on Visualization and Computer Graphics}} \bibinfo{volume}{29}, \bibinfo{number}{1} (\bibinfo{year}{2022}), \bibinfo{pages}{1--11}.
\newblock
\showISSN{1077-2626, 1941-0506, 2160-9306}
\href{https://doi.org/10.1109/TVCG.2022.3209468}{doi:\nolinkurl{10.1109/TVCG.2022.3209468}}


\bibitem[Dibia(2023)]%
        {dibia2023lida}
\bibfield{author}{\bibinfo{person}{Victor Dibia}.} \bibinfo{year}{2023}\natexlab{}.
\newblock \showarticletitle{{LIDA}: A Tool for Automatic Generation of Grammar-Agnostic Visualizations and Infographics using Large Language Models}. In \bibinfo{booktitle}{\emph{Proceedings of the 61st Annual Meeting of the Association for Computational Linguistics (Volume 3: System Demonstrations)}}. \bibinfo{publisher}{Association for Computational Linguistics}, \bibinfo{address}{Toronto, Canada}, \bibinfo{pages}{113--126}.
\newblock
\href{https://doi.org/10.18653/v1/2023.acl-demo.11}{doi:\nolinkurl{10.18653/v1/2023.acl-demo.11}}


\bibitem[Eckelt et~al\mbox{.}(2025)]%
        {eckelt2024loops}
\bibfield{author}{\bibinfo{person}{Klaus Eckelt}, \bibinfo{person}{Kiran Gadhave}, \bibinfo{person}{Alexander Lex}, {and} \bibinfo{person}{Marc Streit}.} \bibinfo{year}{2025}\natexlab{}.
\newblock \showarticletitle{Loops: Leveraging Provenance and Visualization to Support Exploratory Data Analysis in Notebooks}.
\newblock \bibinfo{journal}{\emph{IEEE Transactions on Visualization and Computer Graphics}} \bibinfo{volume}{31}, \bibinfo{number}{1} (\bibinfo{year}{2025}), \bibinfo{pages}{1213--1223}.
\newblock
\href{https://doi.org/10.1109/TVCG.2024.3456186}{doi:\nolinkurl{10.1109/TVCG.2024.3456186}}


\bibitem[Epperson et~al\mbox{.}(2022)]%
        {epperson2022strategiesReuseTeam}
\bibfield{author}{\bibinfo{person}{Will Epperson}, \bibinfo{person}{April~Yi Wang}, \bibinfo{person}{Robert DeLine}, {and} \bibinfo{person}{Steven~M. Drucker}.} \bibinfo{year}{2022}\natexlab{}.
\newblock \showarticletitle{Strategies for reuse and sharing among data scientists in software teams}. In \bibinfo{booktitle}{\emph{Proceedings of the 44th International Conference on Software Engineering: Software Engineering in Practice}} (Pittsburgh, Pennsylvania) \emph{(\bibinfo{series}{ICSE-SEIP '22})}. \bibinfo{publisher}{Association for Computing Machinery}, \bibinfo{address}{New York, NY, USA}, \bibinfo{pages}{243–252}.
\newblock
\showISBNx{9781450392266}
\href{https://doi.org/10.1145/3510457.3513042}{doi:\nolinkurl{10.1145/3510457.3513042}}


\bibitem[Hao et~al\mbox{.}(2024)]%
        {designpatterns2025hao}
\bibfield{author}{\bibinfo{person}{Shan Hao}, \bibinfo{person}{Zezhong Wang}, \bibinfo{person}{Benjamin Bach}, {and} \bibinfo{person}{Larissa Pschetz}.} \bibinfo{year}{2024}\natexlab{}.
\newblock \showarticletitle{Design Patterns for Data-Driven News Articles}. In \bibinfo{booktitle}{\emph{Proceedings of the 2024 CHI Conference on Human Factors in Computing Systems}} (Honolulu, HI, USA) \emph{(\bibinfo{series}{CHI '24})}. \bibinfo{publisher}{Association for Computing Machinery}, \bibinfo{address}{New York, NY, USA}, Article \bibinfo{articleno}{231}, \bibinfo{numpages}{16}~pages.
\newblock
\showISBNx{9798400703300}
\href{https://doi.org/10.1145/3613904.3641916}{doi:\nolinkurl{10.1145/3613904.3641916}}


\bibitem[Hong and Crisan(2023)]%
        {hong2023aithreads}
\bibfield{author}{\bibinfo{person}{Matt-Heun Hong} {and} \bibinfo{person}{Anamaria Crisan}.} \bibinfo{year}{2023}\natexlab{}.
\newblock \showarticletitle{Conversational {AI} Threads for Visualizing Multidimensional Datasets}.
\newblock \bibinfo{journal}{\emph{arXiv preprint arXiv:2311.05590}} (\bibinfo{year}{2023}).
\newblock
\href{https://doi.org/10.48550/arXiv.2311.05590}{doi:\nolinkurl{10.48550/arXiv.2311.05590}}


\bibitem[Huang et~al\mbox{.}(2023)]%
        {huang2023nl2rigel}
\bibfield{author}{\bibinfo{person}{Yanwei Huang}, \bibinfo{person}{Yunfan Zhou}, \bibinfo{person}{Ran Chen}, \bibinfo{person}{Changhao Pan}, \bibinfo{person}{Xinhuan Shu}, \bibinfo{person}{Di Weng}, {and} \bibinfo{person}{Yingcai Wu}.} \bibinfo{year}{2023}\natexlab{}.
\newblock \showarticletitle{Interactive table synthesis with natural language}.
\newblock \bibinfo{journal}{\emph{IEEE Transactions on Visualization and Computer Graphics}} \bibinfo{volume}{30}, \bibinfo{number}{9} (\bibinfo{year}{2023}), \bibinfo{pages}{6130--6145}.
\newblock
\href{https://doi.org/10.1109/TVCG.2023.3329120}{doi:\nolinkurl{10.1109/TVCG.2023.3329120}}


\bibitem[Hunter(2007)]%
        {Hunter2007matplotlib}
\bibfield{author}{\bibinfo{person}{John~D. Hunter}.} \bibinfo{year}{2007}\natexlab{}.
\newblock \showarticletitle{Matplotlib: A 2D graphics environment}.
\newblock \bibinfo{journal}{\emph{Computing in Science \& Engineering}} \bibinfo{volume}{9}, \bibinfo{number}{3} (\bibinfo{year}{2007}), \bibinfo{pages}{90--95}.
\newblock
\href{https://doi.org/10.1109/MCSE.2007.55}{doi:\nolinkurl{10.1109/MCSE.2007.55}}


\bibitem[Islam et~al\mbox{.}(2024)]%
        {islam2024DataNarrativeAutomatedDataDriven}
\bibfield{author}{\bibinfo{person}{Mohammed~Saidul Islam}, \bibinfo{person}{Md~Tahmid~Rahman Laskar}, \bibinfo{person}{Md~Rizwan Parvez}, \bibinfo{person}{Enamul Hoque}, {and} \bibinfo{person}{Shafiq Joty}.} \bibinfo{year}{2024}\natexlab{}.
\newblock \showarticletitle{{D}ata{N}arrative: Automated Data-Driven Storytelling with Visualizations and Texts}. In \bibinfo{booktitle}{\emph{Proceedings of the 2024 Conference on Empirical Methods in Natural Language Processing}}. \bibinfo{publisher}{Association for Computational Linguistics}, \bibinfo{address}{Miami, Florida, USA}, \bibinfo{pages}{19253--19286}.
\newblock
\href{https://doi.org/10.18653/v1/2024.emnlp-main.1073}{doi:\nolinkurl{10.18653/v1/2024.emnlp-main.1073}}


\bibitem[Kang et~al\mbox{.}(2021)]%
        {kang2021toonnote}
\bibfield{author}{\bibinfo{person}{DaYe Kang}, \bibinfo{person}{Tony Ho}, \bibinfo{person}{Nicolai Marquardt}, \bibinfo{person}{Bilge Mutlu}, {and} \bibinfo{person}{Andrea Bianchi}.} \bibinfo{year}{2021}\natexlab{}.
\newblock \showarticletitle{Toonnote: Improving communication in computational notebooks using interactive data comics}. In \bibinfo{booktitle}{\emph{Proceedings of the 2021 CHI Conference on Human Factors in Computing Systems}}. \bibinfo{publisher}{Association for Computing Machinery}, \bibinfo{address}{New York, NY, USA}, \bibinfo{pages}{1--14}.
\newblock
\href{https://doi.org/10.1145/3411764.3445434}{doi:\nolinkurl{10.1145/3411764.3445434}}


\bibitem[Kery and Myers(2017)]%
        {kery2017exploring}
\bibfield{author}{\bibinfo{person}{Mary~Beth Kery} {and} \bibinfo{person}{Brad~A Myers}.} \bibinfo{year}{2017}\natexlab{}.
\newblock \showarticletitle{Exploring exploratory programming}. In \bibinfo{booktitle}{\emph{Proceedings of IEEE Symposium on Visual Languages and Human-Centric Computing}}. IEEE, \bibinfo{pages}{25--29}.
\newblock
\href{https://doi.org/10.1109/VLHCC.2017.8103446}{doi:\nolinkurl{10.1109/VLHCC.2017.8103446}}


\bibitem[Kim et~al\mbox{.}(2019)]%
        {kim2019datatoon}
\bibfield{author}{\bibinfo{person}{Nam~Wook Kim}, \bibinfo{person}{Nathalie Henry~Riche}, \bibinfo{person}{Benjamin Bach}, \bibinfo{person}{Guanpeng Xu}, \bibinfo{person}{Matthew Brehmer}, \bibinfo{person}{Ken Hinckley}, \bibinfo{person}{Michel Pahud}, \bibinfo{person}{Haijun Xia}, \bibinfo{person}{Michael~J McGuffin}, {and} \bibinfo{person}{Hanspeter Pfister}.} \bibinfo{year}{2019}\natexlab{}.
\newblock \showarticletitle{{DataToon}: Drawing dynamic network comics with pen+ touch interaction}. In \bibinfo{booktitle}{\emph{Proceedings of the CHI Conference on Human Factors in Computing Systems}}. \bibinfo{publisher}{Association for Computing Machinery}, \bibinfo{address}{New York, NY, USA}, \bibinfo{pages}{1--12}.
\newblock
\href{https://doi.org/10.1145/3290605.3300335}{doi:\nolinkurl{10.1145/3290605.3300335}}


\bibitem[Koenzen et~al\mbox{.}(2020)]%
        {koenzen2020code}
\bibfield{author}{\bibinfo{person}{Andreas~P Koenzen}, \bibinfo{person}{Neil~A Ernst}, {and} \bibinfo{person}{Margaret-Anne~D Storey}.} \bibinfo{year}{2020}\natexlab{}.
\newblock \showarticletitle{Code duplication and reuse in Jupyter notebooks}. In \bibinfo{booktitle}{\emph{Proceedings of IEEE Symposium on Visual Languages and Human-centric Computing}}. IEEE, \bibinfo{pages}{1--9}.
\newblock
\href{https://doi.org/10.1109/VL/HCC50065.2020.9127202}{doi:\nolinkurl{10.1109/VL/HCC50065.2020.9127202}}


\bibitem[Kojima et~al\mbox{.}(2022)]%
        {kojima2022large}
\bibfield{author}{\bibinfo{person}{Takeshi Kojima}, \bibinfo{person}{Shixiang~(Shane) Gu}, \bibinfo{person}{Machel Reid}, \bibinfo{person}{Yutaka Matsuo}, {and} \bibinfo{person}{Yusuke Iwasawa}.} \bibinfo{year}{2022}\natexlab{}.
\newblock \showarticletitle{Large Language Models are Zero-Shot Reasoners}. In \bibinfo{booktitle}{\emph{Advances in Neural Information Processing Systems}}, Vol.~\bibinfo{volume}{35}. \bibinfo{publisher}{Curran Associates, Inc.}, \bibinfo{pages}{22199--22213}.
\newblock


\bibitem[Li et~al\mbox{.}(2025)]%
        {li2023ai}
\bibfield{author}{\bibinfo{person}{Haotian Li}, \bibinfo{person}{Yun Wang}, \bibinfo{person}{Q.~Vera Liao}, {and} \bibinfo{person}{Huamin Qu}.} \bibinfo{year}{2025}\natexlab{}.
\newblock \showarticletitle{Why Is AI Not a Panacea for Data Workers? An Interview Study on Human-AI Collaboration in Data Storytelling}.
\newblock \bibinfo{journal}{\emph{IEEE Transactions on Visualization and Computer Graphics}} (\bibinfo{year}{2025}), \bibinfo{pages}{1--16}.
\newblock
\href{https://doi.org/10.1109/TVCG.2025.3552017}{doi:\nolinkurl{10.1109/TVCG.2025.3552017}}


\bibitem[Li et~al\mbox{.}(2024)]%
        {li2023wherearewesofar}
\bibfield{author}{\bibinfo{person}{Haotian Li}, \bibinfo{person}{Yun Wang}, {and} \bibinfo{person}{Huamin Qu}.} \bibinfo{year}{2024}\natexlab{}.
\newblock \showarticletitle{Where Are We So Far? Understanding Data Storytelling Tools from the Perspective of Human-AI Collaboration}. In \bibinfo{booktitle}{\emph{Proceedings of the CHI Conference on Human Factors in Computing Systems}}. \bibinfo{publisher}{ACM}, \bibinfo{address}{Honolulu HI USA}, \bibinfo{pages}{1--19}.
\newblock
\showISBNx{979-8-4007-0330-0}
\href{https://doi.org/10.1145/3613904.3642726}{doi:\nolinkurl{10.1145/3613904.3642726}}


\bibitem[Li et~al\mbox{.}(2023)]%
        {li2023edassistant}
\bibfield{author}{\bibinfo{person}{Xingjun Li}, \bibinfo{person}{Yizhi Zhang}, \bibinfo{person}{Justin Leung}, \bibinfo{person}{Chengnian Sun}, {and} \bibinfo{person}{Jian Zhao}.} \bibinfo{year}{2023}\natexlab{}.
\newblock \showarticletitle{{EDAssistant}: Supporting exploratory data analysis in computational notebooks with in situ code search and recommendation}.
\newblock \bibinfo{journal}{\emph{ACM Transactions on Interactive Intelligent Systems}} \bibinfo{volume}{13}, \bibinfo{number}{1} (\bibinfo{year}{2023}), \bibinfo{pages}{1--27}.
\newblock
\href{https://doi.org/10.1145/3545995}{doi:\nolinkurl{10.1145/3545995}}


\bibitem[Lin et~al\mbox{.}(2023)]%
        {lin2023inksight}
\bibfield{author}{\bibinfo{person}{Yanna Lin}, \bibinfo{person}{Haotian Li}, \bibinfo{person}{Leni Yang}, \bibinfo{person}{Aoyu Wu}, {and} \bibinfo{person}{Huamin Qu}.} \bibinfo{year}{2023}\natexlab{}.
\newblock \showarticletitle{{InkSight}: Leveraging sketch interaction for documenting chart findings in computational notebooks}.
\newblock \bibinfo{journal}{\emph{IEEE Transactions on Visualization and Computer Graphics}} \bibinfo{volume}{30}, \bibinfo{number}{1} (\bibinfo{year}{2023}), \bibinfo{pages}{944--954}.
\newblock
\href{https://doi.org/10.1109/TVCG.2023.3327170}{doi:\nolinkurl{10.1109/TVCG.2023.3327170}}


\bibitem[Liu et~al\mbox{.}(2023)]%
        {liu2023refactoring}
\bibfield{author}{\bibinfo{person}{Eric~S. Liu}, \bibinfo{person}{Dylan~A. Lukes}, {and} \bibinfo{person}{William~G. Griswold}.} \bibinfo{year}{2023}\natexlab{}.
\newblock \showarticletitle{Refactoring in Computational Notebooks}.
\newblock \bibinfo{journal}{\emph{ACM Transactions on Software Engineering and Methodology}} \bibinfo{volume}{32}, \bibinfo{number}{3}, Article \bibinfo{articleno}{77} (\bibinfo{year}{2023}), \bibinfo{numpages}{24}~pages.
\newblock
\showISSN{1049-331X}
\href{https://doi.org/10.1145/3576036}{doi:\nolinkurl{10.1145/3576036}}


\bibitem[Lu et~al\mbox{.}(2021)]%
        {lu2021scrollytelling}
\bibfield{author}{\bibinfo{person}{Junhua Lu}, \bibinfo{person}{Wei Chen}, \bibinfo{person}{Hui Ye}, \bibinfo{person}{Jie Wang}, \bibinfo{person}{Honghui Mei}, \bibinfo{person}{Yuhui Gu}, \bibinfo{person}{Yingcai Wu}, \bibinfo{person}{Xiaolong~Luke Zhang}, {and} \bibinfo{person}{Kwan-Liu Ma}.} \bibinfo{year}{2021}\natexlab{}.
\newblock \showarticletitle{Automatic generation of unit visualization-based scrollytelling for impromptu data facts delivery}. In \bibinfo{booktitle}{\emph{Proceedings of IEEE Pacific Visualization Symposium}}. IEEE, \bibinfo{pages}{21--30}.
\newblock
\href{https://doi.org/10.1109/PacificVis52677.2021.00011}{doi:\nolinkurl{10.1109/PacificVis52677.2021.00011}}


\bibitem[Luo et~al\mbox{.}(2025)]%
        {luo2024ferry}
\bibfield{author}{\bibinfo{person}{Zhongsu Luo}, \bibinfo{person}{Kai Xiong}, \bibinfo{person}{Jiajun Zhu}, \bibinfo{person}{Ran Chen}, \bibinfo{person}{Xinhuan Shu}, \bibinfo{person}{Di Weng}, {and} \bibinfo{person}{Yingcai Wu}.} \bibinfo{year}{2025}\natexlab{}.
\newblock \showarticletitle{Ferry: Toward Better Understanding of Input/Output Space for Data Wrangling Scripts}.
\newblock \bibinfo{journal}{\emph{IEEE Transactions on Visualization and Computer Graphics}} \bibinfo{volume}{31}, \bibinfo{number}{1} (\bibinfo{year}{2025}), \bibinfo{pages}{1202--1212}.
\newblock
\href{https://doi.org/10.1109/TVCG.2024.3456328}{doi:\nolinkurl{10.1109/TVCG.2024.3456328}}


\bibitem[Ma et~al\mbox{.}(2023)]%
        {ma2023insightpilot}
\bibfield{author}{\bibinfo{person}{Pingchuan Ma}, \bibinfo{person}{Rui Ding}, \bibinfo{person}{Shuai Wang}, \bibinfo{person}{Shi Han}, {and} \bibinfo{person}{Dongmei Zhang}.} \bibinfo{year}{2023}\natexlab{}.
\newblock \showarticletitle{{I}nsight{P}ilot: An {LLM}-Empowered Automated Data Exploration System}. In \bibinfo{booktitle}{\emph{Proceedings of the 2023 Conference on Empirical Methods in Natural Language Processing: System Demonstrations}}, \bibfield{editor}{\bibinfo{person}{Yansong Feng} {and} \bibinfo{person}{Els Lefever}} (Eds.). \bibinfo{publisher}{Association for Computational Linguistics}, \bibinfo{address}{Singapore}, \bibinfo{pages}{346--352}.
\newblock
\href{https://doi.org/10.18653/v1/2023.emnlp-demo.31}{doi:\nolinkurl{10.18653/v1/2023.emnlp-demo.31}}


\bibitem[Maddigan and Susnjak(2023)]%
        {maddigan2023chat2vis}
\bibfield{author}{\bibinfo{person}{Paula Maddigan} {and} \bibinfo{person}{Teo Susnjak}.} \bibinfo{year}{2023}\natexlab{}.
\newblock \showarticletitle{Chat2VIS: Generating Data Visualizations via Natural Language Using ChatGPT, Codex and GPT-3 Large Language Models}.
\newblock \bibinfo{journal}{\emph{IEEE Access}}  \bibinfo{volume}{11} (\bibinfo{year}{2023}), \bibinfo{pages}{45181--45193}.
\newblock
\href{https://doi.org/10.1109/ACCESS.2023.3274199}{doi:\nolinkurl{10.1109/ACCESS.2023.3274199}}


\bibitem[Masry et~al\mbox{.}(2022)]%
        {masry2022chartqa}
\bibfield{author}{\bibinfo{person}{Ahmed Masry}, \bibinfo{person}{Do~Xuan Long}, \bibinfo{person}{Jia~Qing Tan}, \bibinfo{person}{Shafiq Joty}, {and} \bibinfo{person}{Enamul Hoque}.} \bibinfo{year}{2022}\natexlab{}.
\newblock \showarticletitle{{C}hart{QA}: A Benchmark for Question Answering about Charts with Visual and Logical Reasoning}. In \bibinfo{booktitle}{\emph{Findings of the Association for Computational Linguistics: ACL 2022}}. \bibinfo{publisher}{Association for Computational Linguistics}, \bibinfo{address}{Dublin, Ireland}, \bibinfo{pages}{2263--2279}.
\newblock
\href{https://doi.org/10.18653/v1/2022.findings-acl.177}{doi:\nolinkurl{10.18653/v1/2022.findings-acl.177}}


\bibitem[M{\"o}rth et~al\mbox{.}(2022)]%
        {morth2022scrollyvis}
\bibfield{author}{\bibinfo{person}{Eric M{\"o}rth}, \bibinfo{person}{Stefan Bruckner}, {and} \bibinfo{person}{Noeska~N Smit}.} \bibinfo{year}{2022}\natexlab{}.
\newblock \showarticletitle{{ScrollyVis}: Interactive visual authoring of guided dynamic narratives for scientific scrollytelling}.
\newblock \bibinfo{journal}{\emph{IEEE Transactions on Visualization and Computer Graphics}} \bibinfo{volume}{29}, \bibinfo{number}{12} (\bibinfo{year}{2022}), \bibinfo{pages}{5165--5177}.
\newblock
\href{https://doi.org/10.1109/TVCG.2022.3205769}{doi:\nolinkurl{10.1109/TVCG.2022.3205769}}


\bibitem[nteract(2024)]%
        {papermill}
\bibfield{author}{\bibinfo{person}{nteract}.} \bibinfo{year}{2024}\natexlab{}.
\newblock \bibinfo{title}{papermill: Parameterize, execute, and analyze notebooks}.
\newblock \bibinfo{howpublished}{\url{https://github.com/nteract/papermill}}.
\newblock


\bibitem[ONS({[n.\,d.]})]%
        {ons}
\bibfield{author}{\bibinfo{person}{ONS}.} \bibinfo{year}{[n.\,d.]}\natexlab{}.
\newblock \bibinfo{title}{Office for National Statistics}.
\newblock \bibinfo{howpublished}{\url{https://www.ons.gov.uk/}}.
\newblock


\bibitem[OpenAI(2022)]%
        {chatgpt}
\bibfield{author}{\bibinfo{person}{OpenAI}.} \bibinfo{year}{2022}\natexlab{}.
\newblock \bibinfo{title}{Introducing {ChatGPT}}.
\newblock \bibinfo{howpublished}{\url{https://openai.com/blog/chatgpt}}.
\newblock


\bibitem[OpenAI(2023)]%
        {openai2023gpt4}
\bibfield{author}{\bibinfo{person}{OpenAI}.} \bibinfo{year}{2023}\natexlab{}.
\newblock \showarticletitle{{GPT-4} Technical Report}.
\newblock \bibinfo{journal}{\emph{arXiv preprint arXiv:2303.08774}} (\bibinfo{year}{2023}).
\newblock
\href{https://doi.org/10.48550/arXiv.2303.08774}{doi:\nolinkurl{10.48550/arXiv.2303.08774}}


\bibitem[{Pew Research Center}({[n.\,d.]})]%
        {pewResearchCenter}
\bibfield{author}{\bibinfo{person}{{Pew Research Center}}.} \bibinfo{year}{[n.\,d.]}\natexlab{}.
\newblock \bibinfo{title}{{Pew Research Center}: Nonpartisan, nonadvocacy, public opinion polling and data-driven social science research}.
\newblock \bibinfo{howpublished}{\url{https://www.pewresearch.org/}}.
\newblock


\bibitem[PPIC({[n.\,d.]})]%
        {ppic}
\bibfield{author}{\bibinfo{person}{PPIC}.} \bibinfo{year}{[n.\,d.]}\natexlab{}.
\newblock \bibinfo{title}{Public Policy Institute of California}.
\newblock \bibinfo{howpublished}{\url{https://www.ppic.org/}}.
\newblock


\bibitem[Qian et~al\mbox{.}(2020)]%
        {qian2020retrieve}
\bibfield{author}{\bibinfo{person}{Chunyao Qian}, \bibinfo{person}{Shizhao Sun}, \bibinfo{person}{Weiwei Cui}, \bibinfo{person}{Jian-Guang Lou}, \bibinfo{person}{Haidong Zhang}, {and} \bibinfo{person}{Dongmei Zhang}.} \bibinfo{year}{2020}\natexlab{}.
\newblock \showarticletitle{Retrieve-then-adapt: Example-based automatic generation for proportion-related infographics}.
\newblock \bibinfo{journal}{\emph{IEEE Transactions on Visualization and Computer Graphics}} \bibinfo{volume}{27}, \bibinfo{number}{2} (\bibinfo{year}{2020}), \bibinfo{pages}{443--452}.
\newblock
\href{https://doi.org/10.1109/TVCG.2020.3030448}{doi:\nolinkurl{10.1109/TVCG.2020.3030448}}


\bibitem[Raghunandan et~al\mbox{.}(2023)]%
        {raghunandan2023codeEvolution}
\bibfield{author}{\bibinfo{person}{Deepthi Raghunandan}, \bibinfo{person}{Aayushi Roy}, \bibinfo{person}{Shenzhi Shi}, \bibinfo{person}{Niklas Elmqvist}, {and} \bibinfo{person}{Leilani Battle}.} \bibinfo{year}{2023}\natexlab{}.
\newblock \showarticletitle{Code Code Evolution: Understanding How People Change Data Science Notebooks Over Time}. In \bibinfo{booktitle}{\emph{Proceedings of the 2023 CHI Conference on Human Factors in Computing Systems}} (Hamburg, Germany) \emph{(\bibinfo{series}{CHI '23})}. \bibinfo{publisher}{Association for Computing Machinery}, \bibinfo{address}{New York, NY, USA}, Article \bibinfo{articleno}{863}, \bibinfo{numpages}{12}~pages.
\newblock
\showISBNx{9781450394215}
\href{https://doi.org/10.1145/3544548.3580997}{doi:\nolinkurl{10.1145/3544548.3580997}}


\bibitem[Ritta et~al\mbox{.}(2022)]%
        {ritta2022reusingMyOwnCode}
\bibfield{author}{\bibinfo{person}{Natanon Ritta}, \bibinfo{person}{Tasha Settewong}, \bibinfo{person}{Raula~Gaikovina Kula}, \bibinfo{person}{Chaiyong Ragkhitwetsagul}, \bibinfo{person}{Thanwadee Sunetnanta}, {and} \bibinfo{person}{Kenichi Matsumoto}.} \bibinfo{year}{2022}\natexlab{}.
\newblock \showarticletitle{Reusing My Own Code: Preliminary Results for Competitive Coding in Jupyter Notebooks}. In \bibinfo{booktitle}{\emph{Proceedings of Asia-Pacific Software Engineering Conference}}. IEEE, \bibinfo{pages}{457--461}.
\newblock
\href{https://doi.org/10.1109/APSEC57359.2022.00062}{doi:\nolinkurl{10.1109/APSEC57359.2022.00062}}


\bibitem[Segel and Heer(2010)]%
        {segel2010narrative}
\bibfield{author}{\bibinfo{person}{Edward Segel} {and} \bibinfo{person}{Jeffrey Heer}.} \bibinfo{year}{2010}\natexlab{}.
\newblock \showarticletitle{Narrative visualization: Telling stories with data}.
\newblock \bibinfo{journal}{\emph{IEEE Transactions on Visualization and Computer Graphics}} \bibinfo{volume}{16}, \bibinfo{number}{6} (\bibinfo{year}{2010}), \bibinfo{pages}{1139--1148}.
\newblock
\href{https://doi.org/10.1109/TVCG.2010.179}{doi:\nolinkurl{10.1109/TVCG.2010.179}}


\bibitem[Shen et~al\mbox{.}(2024)]%
        {FromDatatoStory}
\bibfield{author}{\bibinfo{person}{Leixian Shen}, \bibinfo{person}{Haotian Li}, \bibinfo{person}{Yun Wang}, {and} \bibinfo{person}{Huamin Qu}.} \bibinfo{year}{2024}\natexlab{}.
\newblock \showarticletitle{From Data to Story: Towards Automatic Animated Data Video Creation with LLM-Based Multi-Agent Systems}. In \bibinfo{booktitle}{\emph{2024 IEEE VIS Workshop on Data Storytelling in an Era of Generative AI (GEN4DS)}}. \bibinfo{pages}{20--27}.
\newblock
\href{https://doi.org/10.1109/GEN4DS63889.2024.00008}{doi:\nolinkurl{10.1109/GEN4DS63889.2024.00008}}


\bibitem[Shi et~al\mbox{.}(2020)]%
        {shi2020calliope}
\bibfield{author}{\bibinfo{person}{Danqing Shi}, \bibinfo{person}{Xinyue Xu}, \bibinfo{person}{Fuling Sun}, \bibinfo{person}{Yang Shi}, {and} \bibinfo{person}{Nan Cao}.} \bibinfo{year}{2020}\natexlab{}.
\newblock \showarticletitle{Calliope: Automatic visual data story generation from a spreadsheet}.
\newblock \bibinfo{journal}{\emph{IEEE Transactions on Visualization and Computer Graphics}} \bibinfo{volume}{27}, \bibinfo{number}{2} (\bibinfo{year}{2020}), \bibinfo{pages}{453--463}.
\newblock
\href{https://doi.org/10.1109/TVCG.2020.3030403}{doi:\nolinkurl{10.1109/TVCG.2020.3030403}}


\bibitem[Shin et~al\mbox{.}(2022)]%
        {shin2022roslingifier}
\bibfield{author}{\bibinfo{person}{Minjeong Shin}, \bibinfo{person}{Joohee Kim}, \bibinfo{person}{Yunha Han}, \bibinfo{person}{Lexing Xie}, \bibinfo{person}{Mitchell Whitelaw}, \bibinfo{person}{Bum~Chul Kwon}, \bibinfo{person}{Sungahn Ko}, {and} \bibinfo{person}{Niklas Elmqvist}.} \bibinfo{year}{2022}\natexlab{}.
\newblock \showarticletitle{Roslingifier: Semi-automated storytelling for animated scatterplots}.
\newblock \bibinfo{journal}{\emph{IEEE Transactions on Visualization and Computer Graphics}} \bibinfo{volume}{29}, \bibinfo{number}{6} (\bibinfo{year}{2022}), \bibinfo{pages}{2980--2995}.
\newblock
\href{https://doi.org/10.1109/TVCG.2022.3146329}{doi:\nolinkurl{10.1109/TVCG.2022.3146329}}


\bibitem[Subramanian et~al\mbox{.}(2020)]%
        {subramanian2020tractus}
\bibfield{author}{\bibinfo{person}{Krishna Subramanian}, \bibinfo{person}{Johannes Maas}, {and} \bibinfo{person}{Jan Borchers}.} \bibinfo{year}{2020}\natexlab{}.
\newblock \showarticletitle{{TRACTUS}: Understanding and supporting source code experimentation in hypothesis-driven data science}. In \bibinfo{booktitle}{\emph{Proceedings of the CHI Conference on Human Factors in Computing Systems}}. \bibinfo{publisher}{Association for Computing Machinery}, \bibinfo{address}{New York, NY, USA}, \bibinfo{pages}{1--12}.
\newblock
\href{https://doi.org/10.1145/3313831.3376764}{doi:\nolinkurl{10.1145/3313831.3376764}}


\bibitem[Suh et~al\mbox{.}(2023)]%
        {suh2023Sensecape}
\bibfield{author}{\bibinfo{person}{Sangho Suh}, \bibinfo{person}{Bryan Min}, \bibinfo{person}{Srishti Palani}, {and} \bibinfo{person}{Haijun Xia}.} \bibinfo{year}{2023}\natexlab{}.
\newblock \showarticletitle{Sensecape: Enabling Multilevel Exploration and Sensemaking with Large Language Models}. In \bibinfo{booktitle}{\emph{Proceedings of the 36th Annual ACM Symposium on User Interface Software and Technology}}. \bibinfo{publisher}{Association for Computing Machinery}, \bibinfo{address}{New York, NY, USA}, \bibinfo{numpages}{18}~pages.
\newblock
\href{https://doi.org/10.1145/3586183.3606756}{doi:\nolinkurl{10.1145/3586183.3606756}}


\bibitem[Sultanum and Srinivasan(2023)]%
        {sultanum2023datatales}
\bibfield{author}{\bibinfo{person}{Nicole Sultanum} {and} \bibinfo{person}{Arjun Srinivasan}.} \bibinfo{year}{2023}\natexlab{}.
\newblock \showarticletitle{DATATALES: Investigating the use of large language models for authoring data-driven articles}. In \bibinfo{booktitle}{\emph{IEEE Visualization and Visual Analytics (VIS)}}. IEEE, \bibinfo{pages}{231--235}.
\newblock
\href{https://doi.org/10.1109/VIS54172.2023.00055}{doi:\nolinkurl{10.1109/VIS54172.2023.00055}}


\bibitem[Sun et~al\mbox{.}(2022)]%
        {sun2022erato}
\bibfield{author}{\bibinfo{person}{Mengdi Sun}, \bibinfo{person}{Ligan Cai}, \bibinfo{person}{Weiwei Cui}, \bibinfo{person}{Yanqiu Wu}, \bibinfo{person}{Yang Shi}, {and} \bibinfo{person}{Nan Cao}.} \bibinfo{year}{2022}\natexlab{}.
\newblock \showarticletitle{Erato: Cooperative data story editing via fact interpolation}.
\newblock \bibinfo{journal}{\emph{IEEE Transactions on Visualization and Computer Graphics}} \bibinfo{volume}{29}, \bibinfo{number}{1} (\bibinfo{year}{2022}), \bibinfo{pages}{983--993}.
\newblock
\href{https://doi.org/10.1109/TVCG.2022.3209428}{doi:\nolinkurl{10.1109/TVCG.2022.3209428}}


\bibitem[Tian et~al\mbox{.}(2025)]%
        {tian2024chartgpt}
\bibfield{author}{\bibinfo{person}{Yuan Tian}, \bibinfo{person}{Weiwei Cui}, \bibinfo{person}{Dazhen Deng}, \bibinfo{person}{Xinjing Yi}, \bibinfo{person}{Yurun Yang}, \bibinfo{person}{Haidong Zhang}, {and} \bibinfo{person}{Yingcai Wu}.} \bibinfo{year}{2025}\natexlab{}.
\newblock \showarticletitle{{ChartGPT}: Leveraging LLMs to Generate Charts From Abstract Natural Language}.
\newblock \bibinfo{journal}{\emph{IEEE Transactions on Visualization and Computer Graphics}} \bibinfo{volume}{31}, \bibinfo{number}{3} (\bibinfo{year}{2025}), \bibinfo{pages}{1731--1745}.
\newblock
\href{https://doi.org/10.1109/TVCG.2024.3368621}{doi:\nolinkurl{10.1109/TVCG.2024.3368621}}


\bibitem[Wang et~al\mbox{.}(2022)]%
        {wang2022documentation}
\bibfield{author}{\bibinfo{person}{April~Yi Wang}, \bibinfo{person}{Dakuo Wang}, \bibinfo{person}{Jaimie Drozdal}, \bibinfo{person}{Michael Muller}, \bibinfo{person}{Soya Park}, \bibinfo{person}{Justin~D Weisz}, \bibinfo{person}{Xuye Liu}, \bibinfo{person}{Lingfei Wu}, {and} \bibinfo{person}{Casey Dugan}.} \bibinfo{year}{2022}\natexlab{}.
\newblock \showarticletitle{Documentation matters: Human-centered AI system to assist data science code documentation in computational notebooks}.
\newblock \bibinfo{journal}{\emph{ACM Transactions on Computer-Human Interaction}} \bibinfo{volume}{29}, \bibinfo{number}{2} (\bibinfo{year}{2022}), \bibinfo{pages}{1--33}.
\newblock
\href{https://doi.org/10.1145/3489465}{doi:\nolinkurl{10.1145/3489465}}


\bibitem[Wang et~al\mbox{.}(2023)]%
        {wang2023dataFormulator}
\bibfield{author}{\bibinfo{person}{Chenglong Wang}, \bibinfo{person}{John Thompson}, {and} \bibinfo{person}{Bongshin Lee}.} \bibinfo{year}{2023}\natexlab{}.
\newblock \showarticletitle{Data Formulator: {AI}-powered concept-driven visualization authoring}.
\newblock \bibinfo{journal}{\emph{IEEE Transactions on Visualization and Computer Graphics}} \bibinfo{volume}{30}, \bibinfo{number}{1} (\bibinfo{year}{2023}), \bibinfo{pages}{1128--1138}.
\newblock
\href{https://doi.org/10.1109/TVCG.2023.3326585}{doi:\nolinkurl{10.1109/TVCG.2023.3326585}}


\bibitem[Wang et~al\mbox{.}(2019)]%
        {wang2019datashot}
\bibfield{author}{\bibinfo{person}{Yun Wang}, \bibinfo{person}{Zhida Sun}, \bibinfo{person}{Haidong Zhang}, \bibinfo{person}{Weiwei Cui}, \bibinfo{person}{Ke Xu}, \bibinfo{person}{Xiaojuan Ma}, {and} \bibinfo{person}{Dongmei Zhang}.} \bibinfo{year}{2019}\natexlab{}.
\newblock \showarticletitle{{DataShot}: Automatic generation of fact sheets from tabular data}.
\newblock \bibinfo{journal}{\emph{IEEE Transactions on Visualization and Computer Graphics}} \bibinfo{volume}{26}, \bibinfo{number}{1} (\bibinfo{year}{2019}), \bibinfo{pages}{895--905}.
\newblock
\href{https://doi.org/10.1109/TVCG.2019.2934398}{doi:\nolinkurl{10.1109/TVCG.2019.2934398}}


\bibitem[Waskom(2021)]%
        {Waskom2021seaborn}
\bibfield{author}{\bibinfo{person}{Michael~L. Waskom}.} \bibinfo{year}{2021}\natexlab{}.
\newblock \showarticletitle{seaborn: statistical data visualization}.
\newblock \bibinfo{journal}{\emph{Journal of Open Source Software}} \bibinfo{volume}{6}, \bibinfo{number}{60} (\bibinfo{year}{2021}), \bibinfo{pages}{3021}.
\newblock
\href{https://doi.org/10.21105/joss.03021}{doi:\nolinkurl{10.21105/joss.03021}}


\bibitem[Wongsuphasawat et~al\mbox{.}(2017)]%
        {wongsuphasawat2017voyager}
\bibfield{author}{\bibinfo{person}{Kanit Wongsuphasawat}, \bibinfo{person}{Zening Qu}, \bibinfo{person}{Dominik Moritz}, \bibinfo{person}{Riley Chang}, \bibinfo{person}{Felix Ouk}, \bibinfo{person}{Anushka Anand}, \bibinfo{person}{Jock Mackinlay}, \bibinfo{person}{Bill Howe}, {and} \bibinfo{person}{Jeffrey Heer}.} \bibinfo{year}{2017}\natexlab{}.
\newblock \showarticletitle{Voyager 2: Augmenting visual analysis with partial view specifications}. In \bibinfo{booktitle}{\emph{Proceedings of the CHI Conference on Human Factors in Computing Systems}}. \bibinfo{publisher}{Association for Computing Machinery}, \bibinfo{address}{New York, NY, USA}, \bibinfo{pages}{2648--2659}.
\newblock
\href{https://doi.org/10.1145/3025453.3025768}{doi:\nolinkurl{10.1145/3025453.3025768}}


\bibitem[Wu et~al\mbox{.}(2024)]%
        {Socrates}
\bibfield{author}{\bibinfo{person}{Guande Wu}, \bibinfo{person}{Shunan Guo}, \bibinfo{person}{Jane Hoffswell}, \bibinfo{person}{Gromit Yeuk-Yin Chan}, \bibinfo{person}{Ryan~A. Rossi}, {and} \bibinfo{person}{Eunyee Koh}.} \bibinfo{year}{2024}\natexlab{}.
\newblock \showarticletitle{Socrates: Data Story Generation via Adaptive Machine-Guided Elicitation of User Feedback}.
\newblock \bibinfo{journal}{\emph{IEEE Transactions on Visualization and Computer Graphics}} \bibinfo{volume}{30}, \bibinfo{number}{1} (\bibinfo{year}{2024}), \bibinfo{pages}{131--141}.
\newblock
\href{https://doi.org/10.1109/TVCG.2023.3327363}{doi:\nolinkurl{10.1109/TVCG.2023.3327363}}


\bibitem[Xie et~al\mbox{.}(2024)]%
        {HAIChart}
\bibfield{author}{\bibinfo{person}{Yupeng Xie}, \bibinfo{person}{Yuyu Luo}, \bibinfo{person}{Guoliang Li}, {and} \bibinfo{person}{Nan Tang}.} \bibinfo{year}{2024}\natexlab{}.
\newblock \showarticletitle{HAIChart: Human and AI Paired Visualization System}.
\newblock \bibinfo{journal}{\emph{Proc. VLDB Endow.}} \bibinfo{volume}{17}, \bibinfo{number}{11} (\bibinfo{year}{2024}), \bibinfo{pages}{3178–3191}.
\newblock
\showISSN{2150-8097}
\href{https://doi.org/10.14778/3681954.3681992}{doi:\nolinkurl{10.14778/3681954.3681992}}


\bibitem[Ying et~al\mbox{.}(2022)]%
        {ying2022metaglyph}
\bibfield{author}{\bibinfo{person}{Lu Ying}, \bibinfo{person}{Xinhuan Shu}, \bibinfo{person}{Dazhen Deng}, \bibinfo{person}{Yuchen Yang}, \bibinfo{person}{Tan Tang}, \bibinfo{person}{Lingyun Yu}, {and} \bibinfo{person}{Yingcai Wu}.} \bibinfo{year}{2022}\natexlab{}.
\newblock \showarticletitle{{MetaGlyph}: Automatic generation of metaphoric glyph-based visualization}.
\newblock \bibinfo{journal}{\emph{IEEE Transactions on Visualization and Computer Graphics}} \bibinfo{volume}{29}, \bibinfo{number}{1} (\bibinfo{year}{2022}), \bibinfo{pages}{331--341}.
\newblock
\href{https://doi.org/10.1109/TVCG.2022.3209447}{doi:\nolinkurl{10.1109/TVCG.2022.3209447}}


\bibitem[Ying et~al\mbox{.}(2025)]%
        {ying2023livecharts}
\bibfield{author}{\bibinfo{person}{Lu Ying}, \bibinfo{person}{Yun Wang}, \bibinfo{person}{Haotian Li}, \bibinfo{person}{Shuguang Dou}, \bibinfo{person}{Haidong Zhang}, \bibinfo{person}{Xinyang Jiang}, \bibinfo{person}{Huamin Qu}, {and} \bibinfo{person}{Yingcai Wu}.} \bibinfo{year}{2025}\natexlab{}.
\newblock \showarticletitle{Reviving Static Charts Into Live Charts}.
\newblock \bibinfo{journal}{\emph{IEEE Transactions on Visualization and Computer Graphics}} \bibinfo{volume}{31}, \bibinfo{number}{8} (\bibinfo{year}{2025}), \bibinfo{pages}{4314--4328}.
\newblock
\href{https://doi.org/10.1109/TVCG.2024.3397004}{doi:\nolinkurl{10.1109/TVCG.2024.3397004}}


\bibitem[YouGov({[n.\,d.]})]%
        {yougov}
\bibfield{author}{\bibinfo{person}{YouGov}.} \bibinfo{year}{[n.\,d.]}\natexlab{}.
\newblock \bibinfo{title}{{YouGov}: What the world thinks}.
\newblock \bibinfo{howpublished}{\url{https://yougov.co.uk/}}.
\newblock


\bibitem[Zhao et~al\mbox{.}(2020)]%
        {zhao2020chartseer}
\bibfield{author}{\bibinfo{person}{Jian Zhao}, \bibinfo{person}{Mingming Fan}, {and} \bibinfo{person}{Mi Feng}.} \bibinfo{year}{2020}\natexlab{}.
\newblock \showarticletitle{{ChartSeer}: Interactive steering exploratory visual analysis with machine intelligence}.
\newblock \bibinfo{journal}{\emph{IEEE Transactions on Visualization and Computer Graphics}} \bibinfo{volume}{28}, \bibinfo{number}{3} (\bibinfo{year}{2020}), \bibinfo{pages}{1500--1513}.
\newblock
\href{https://doi.org/10.1109/TVCG.2020.3018724}{doi:\nolinkurl{10.1109/TVCG.2020.3018724}}


\bibitem[Zhao et~al\mbox{.}(2024)]%
        {zhaoLightVALightweightVisual2024}
\bibfield{author}{\bibinfo{person}{Yuheng Zhao}, \bibinfo{person}{Junjie Wang}, \bibinfo{person}{Linbin Xiang}, \bibinfo{person}{Xiaowen Zhang}, \bibinfo{person}{Zifei Guo}, \bibinfo{person}{Cagatay Turkay}, \bibinfo{person}{Yu Zhang}, {and} \bibinfo{person}{Siming Chen}.} \bibinfo{year}{2024}\natexlab{}.
\newblock \showarticletitle{LightVA: Lightweight Visual Analytics with LLM Agent-Based Task Planning and Execution}.
\newblock \bibinfo{journal}{\emph{IEEE Transactions on Visualization and Computer Graphics}} (\bibinfo{year}{2024}), \bibinfo{pages}{1--13}.
\newblock
\href{https://doi.org/10.1109/TVCG.2024.3496112}{doi:\nolinkurl{10.1109/TVCG.2024.3496112}}


\end{thebibliography}

\end{document}